\documentclass[twocolumn,amsmath,trackchanges]{aastex702}

\newcommand{\A}{{\sc Athena} }
\newcommand{\M}{$M_{\rm exp}$}
\newcommand{\D}{$D_{\rm exp}$}

\newcommand{\Fiducial}{\texttt{Tau1\_Pi05\_Z5\_G1}}
\newcommand{\LowR}{\Fiducial\texttt{-512}}
\newcommand{\UltraLowR}{\Fiducial\texttt{-256}}
\newcommand{\LowZ}{\texttt{Tau1\_Pi05\_Z2\_G1}}
\newcommand{\LowG}{\texttt{Tau1\_Pi05\_Z5\_G05}}
\newcommand{\HighP}{\texttt{Tau1\_Pi1\_Z5\_G1}}
\newcommand{\LowT}{\texttt{Tau01\_Pi05\_Z5\_G1}}
\newcommand{\Gole}{\texttt{Tau3\_Pi05\_Z2\_G1}}

\usepackage{amsmath}
\usepackage{tabularx}
\usepackage{multirow}
\usepackage{soul}
\usepackage{CJKutf8}
\usepackage[utf8]{inputenc}

\newcommand{\bmath}[1]{\mbox{\boldmath{$#1$}}}

\definecolor{darkgreen}{HTML}{006400}
\definecolor{huntergreen}{HTML}{355E3B}
\definecolor{forestgreen}{HTML}{2B8843}
\definecolor{verydarkgreen}{HTML}{004225}
\definecolor{phthalogreen}{HTML}{123524}

\newcommand{\dc}[1]{{\color{forestgreen}#1}}

\begin{document}

\title{Closing the Gap: A Test of the Streaming Instability Using Simulations and Observations of the Cold-Classical Kuiper Belt Size Distribution}

\author[0000-0001-6494-6098]{Weston Hall}
\affiliation{Department of Physics and Astronomy, Iowa State University, Ames, Iowa, USA}
\email{wmhall@iastate.edu}

\author[0000-0002-3771-8054]{Jacob B. Simon}
\affiliation{Department of Physics and Astronomy, Iowa State University, Ames, Iowa, USA}
\email{jbsimon@iastate.edu}

\author[0000-0003-2719-6640]{Jeonghoon Lim (\begin{CJK}{UTF8}{mj}임정훈\end{CJK})}
\affiliation{Department of Physics and Astronomy, Iowa State University, Ames, Iowa, USA}
\affiliation{Nevada Center for Astrophysics, University of Nevada, Las Vegas, Box 454002, Las Vegas, NV 89154, USA}
\affiliation{Department of Physics and Astronomy, University of Nevada, Las Vegas, Box 454002, Las Vegas, NV 89154, USA}
\email{jay.lim1@unlv.edu}

\author[0000-0001-6259-3575]{Daniel Carrera}
\affiliation{Department of Physics and Astronomy, Iowa State University, Ames, Iowa, USA}
\affiliation{Department of Astronomy, New Mexico State University, Las Cruces, New Mexico, USA}
\email{carrera4@nmsu.edu}


\author{Matthew Small}
\affiliation{Department of Physics and Astronomy, Iowa State University, Ames, Iowa, USA}
\email{matthewsmall1995@gmail.com}

\begin{abstract}

How planetesimals form is a major open question in studies of planet formation, as the growth of smaller solids must overcome barriers such as radial drift, fragmentation, and bouncing. 
The leading model is that planetesimals form by the gravitational collapse of dense particle clumps that have been aerodynamically collected via a process known as the streaming instability (SI). While the SI has passed several observational tests, a critical test remains: a successful comparison with the size distribution of Solar System planetesimals. In this work, we conducted a series of SI simulations for a range of input parameters and combine a more careful accounting of initial planetesimal mass with more advanced statistical techniques compared to previous studies in order to compare the simulation outcomes against observations of the most primordial population in the solar system: the Cold Classical Kuiper Belt (CCKB).
Using both cumulative distribution function (CDF) model parameters and distribution comparison methods like the Kolomogorov-Smirnov test, we show that, when planetesimal masses are measured at formation and analyzed with modern statistical methods, the SI-predicted size distributions can quantitatively reproduce the high-mass end of the observed CCKB size distribution, though the extent of the agreement depends on the physical parameters chosen, providing further support for the SI theory of planetesimal formation.

\end{abstract}



\section{Introduction} \label{sec:intro}

Planet formation is a process that spans more than 13 orders of magnitude in spatial scale. A crucial step along the way from $\lesssim \mu {\rm m}$ grains to $\sim 10^4$-$10^5~{\rm km}$ planets is the formation of planetesimals, objects roughly $1$-$100$~km in size that are held together with their own gravity. The formation of planetesimals faces several barriers, including radial drift, where orbiting particles lose angular momentum due to gas drag, causing them to move inward on potentially very short timescales \citep{1977MNRAS.180...57W}, as well as collisional barriers such as bouncing and fragmentation of larger solids into smaller grains \citep{2008ARA&A..46...21B,2010A&A...513A..57Z,2010AREPS..38..493C,2014prpl.conf..547J,2018SSRv..214...52B}. 

Over the past two decades, the streaming instability (SI) has emerged as a promising candidate to overcome these barriers. The SI arises from solid particles and gas in protoplanetary disks exchanging angular momentum through aerodynamic drag \citep{2005ApJ...620..459Y,Youdin_Johansen_2007}; this results in the formation of dense filamentary structures of solid particles \citep{2007ApJ...662..627J,Johansen_2009, Bai_Stone_2010a,2014ApJ...792...86Y,Abod_2019}. Within these filaments, particle clumps that have local overdensities exceeding the Roche density gravitationally collapse to produce bodies at the $\sim$ 100 km scale \citep[see e.g.][]{ 2015SciA....1E0109J,Simon_2016}, which roughly agrees with many of the largest planetesimals in the Solar System.

Adding to this preliminary evidence favoring the SI model, detailed studies of individual gravitationally bound clumps produced by the SI demonstrate that this mechanism favors the formation of binary systems with angular momentum distributions strongly consistent with those observed in the Cold Classical Kuiper Belt (CCKB), including the predominance of prograde rotation and a retrograde fraction of roughly 20\% \citep{Nesvorny_binary, 2021PSJ.....2...27N}. However, more recent work \citep{2022A&A...663A.164V} has demonstrated that the preference for prograde binaries may simply be the result of gravitational collapse of a pebble cloud orbiting in a Keplerian potential, though they did not show that the 80\%/20\% split of prograde/retrograde is an outcome of this collapse. Thus, while the properties of KBO binaries may provide strong support of the SI model, the remaining uncertainties motivate us to find alternative tests of this mechanism. 

One such additional observational test is a comparison between the observed and predicted initial size distributions of planetesimals in the Solar System. While the SI seems to favor planetesimals being born big, in agreement with the size distribution of asteroids in our solar system \citep{2009Icar..204..558M}, more work needs to be done to verify that the size distributions of planetesimal populations is in agreement with SI models. In particular, more robust statistical comparisons between observed and simulated planetesimal populations need to be carried out over a range of planetesimal size scales.

In order to make progress along this front, several previous studies have focused on quantifying the planetesimal initial mass function (IMF). They found relatively top-heavy distributions with a large fraction of the mass contained in the largest objects \citep{2015SciA....1E0109J, Simon_2016,Schafer2017}. \citet{Simon_2017} further suggested that the shape (in that work, they fit a simple power law to their IMFs and compared the power law indicies) of the IMF was largely insensitive to particle stopping time, interpreting this as evidence for a degree of universality in SI-driven collapse; however subsequent studies expanded both the accessible mass range and the explored parameter space, enabled in part by improved clump-identification algorithms, demonstrating that the IMF can vary in both functional form and best-fit parameters \citep{Li_2019}. Despite the progress made by this early work, however, agreement between observed and predicted size distributions has remained elusive (see the review by \citealt{Jakes_Review_chapter}, as well as discussion in \citealt{Bernstein2026}).



On the observational front, the magnitude (and thus size) distribution of the Cold Classical Kuiper Belt objects (CCKBOs) has been studied extensively (e.g., \citealt{2014ApJ...782..100F}). The CCKBOs are the most pristine population of objects observationally accessible, having undergone little to no collisional evolution (but see \citealt{2012ApJ...744..139P,2016MNRAS.455.4487B} for a discussion of the collisional evolution of 10-30\% of CCKBOs that may have occurred), and containing loosely bound binary pairs implying \textit{in situ} formation (\citealt{2010ApJ...722L.204P}; but see \citealt{2025NatAs...9...75C}). At low-magnitudes (large sizes), the cumulative size distribution has an exponential taper \citep{Kavelaars}, and dimmer objects have a size distribution described by a power law \citep{Napier, Eduardo2026,Bernstein2026}.

Most recently, \citet{Bernstein2026} note that the observed exponential taper in size is much gentler than previous SI-driven IMF works suggest, further solidifying earlier concerns of disagreement between observed and predicted size distributions \citep{Jakes_Review_chapter}.
In this work, we attempt to reconcile these differences via several methodological improvements. One noted issue  in aforementioned SI-driven IMF works that these studies typically rely on snapshot-based identification of bound clumps, which can bias inferred masses due to processes such as mergers, tidal stripping, and continued accretion after formation.  Therefore, in addition to varying key physical parameters and employing modern clump-identification methods, we crucially incorporate clump tracking to measure planetesimal masses at the time of formation \citep[c.f.][]{Gole_2020}. We then employ the use of a robust statistical framework to enable direct and rigorous comparisons between predicted and observed size distributions. We find that, depending on the values of the key physical parameters that control the outcome of the SI, our simulations can quantitatively reproduce the high-mass end of the observed CCKB distribution.

This paper is organized as follows. In Section~\ref{sec:methods}, we describe our numerical algorithm in detail, including initial conditions and parameter choices, how to identify planetesimals and construct a predicted mass (and thus size) distribution, the disk model to convert from code to physical units, and the statistical tools we use to characterize the distribution. We present our results for the distribution characteristics and observational comparisons in Section~\ref{sec:results}. In Section~\ref{sec:uncertainties}, we discuss potential impacts of our assumptions on the predicted distribution, along with the associated uncertainties. We summarize and conclude in Section~\ref{sec:conclusions}. 

\section{Methods} \label{sec:methods}

As we describe in the following subsections, we simulated several local, co-rotating disk shearing boxes containing gas and solid particles. The SI clumps the particles until filament structures of solids are formed, after which particle self-gravity is turned on. Planetesimal masses and physical properties are captured at the moment of formation, as close to an actual {\it initial} mass function as possible in our simulations. Readers familiar with the shearing box and IMF analysis methods can skip to Section~\ref{sec:model} for our approach to IMF fitting, and to Section~\ref{sec:results} for our main results.






\subsection{Numerical Method} \label{sec:SI}

We use the {\sc Athena} \citep{2008ApJS..178..137S} code to simulate a small, vertically stratified section of a protoplanetary disk within the local  shearing box approximation \citep{1995ApJ...440..742H, 2010ApJS..189..142S}. For algorithmic choices in {\sc Athena}, we employ the corner transport upwind integrator \citep[CTU;][]{CTU}, the piecewise parabolic method for third order spatial reconstruction \citep[PPM;][]{PPM}, and the Harten-Lax-van Leer with contact Reimann solver for flux quantities \citep[HLLC;][]{HLLC}. 

The shearing box approximation is justified by the small size of the box compared to the large distance from the star, and maps the (cylindrical) disk coordinates ($R,\phi,z'$) onto a local Cartesian system ($x,y,z$) \citep{1965MNRAS.130..125G}. The local box is centered at a radius $R_0$ from the central star in the midplane where:
$x = R - R_0$, $y = R_0 \phi$, $z = z'$ and the Keplerian frequency and velocity are $\Omega$ and $v_K=\Omega R_0$, respectively. We employ orbital advection, in which the background Keplerian shear flow is analytically separated from the dynamical velocity field \citep{2010ApJS..189..142S}. In this approach, the total velocity is decomposed into a prescribed linear shear and a residual velocity (see below), the latter of which is evolved by the hydrodynamic solver. The background shear is advanced as a separate advection step, allowing for larger timesteps and reduced numerical diffusion while preserving the correct epicyclic dynamics of the shearing box. All velocities reported in this work correspond to the residual velocity $\bmath{u}'$. The gas velocity transformation is 

\begin{equation}
    \bmath{u}' = \bmath{u} + \left(\frac{3}{2}\Omega x\right)\hat{y}.
\end{equation}


In this domain, for an isothermal fluid of density $\rho_g$, velocity $\bmath{u}$, and pressure $P$, {\sc Athena} solves the equations of gas dynamics:
\begin{equation}
    \frac{\partial \rho_g }{\partial t} + \nabla \cdot \left( \rho_g \bmath{u} \right) = 0
\end{equation}
\begin{equation}
\begin{split}
    &\frac{\partial \rho_g \bmath{u}}{\partial t} + \nabla \cdot \left(\rho_g \bmath{uu} + P\textbf{I}\right) = \\
    &\rho_g \left( 2\bmath{u} \times \bmath{\Omega} + 3 \Omega^2\bmath{x} - \Omega^2 \bmath{z}\right) + \rho_p\frac{\bar{\bmath{v}}-\bmath{u}}{t_{\rm stop}},
\end{split}
\end{equation}
\begin{equation}
    P=\rho_gc_s^2
\end{equation}
where $\rho_p$ and $\bar{\bmath{v}}$ are the average density and velocity of particles in a hydrodynamic grid cell, $\textbf{I}$ is the identity matrix, $\bmath{\Omega}=\Omega\hat{z}$, $t_{\rm stop}$ is the timescale over which particles lose a significant fraction of their momentum due to gas drag, and $c_s$ is the sound speed. 

For the motion of each particle in {\sc Athena}, we use the semi-implicit time integrator and the triangular-shaped cloud (TSC) method for interpolating particle momentum to nearby grid cells to account for the backreaction on the gas \citep{Bai_Stone_2010a}. For the $i$-th particle with velocity $\bmath{v}_i$, the equation of motion is
\begin{equation}
\label{eq:3}
\begin{split}
    \frac{d \bmath{v}_i}{d t} = 
    & 2 \bmath{v}_i \times \bmath{\Omega} + 3\Omega^2\bmath{x}_i - \Omega^2 \bmath{z}_i \\
    &- \frac{\bmath{v}_i - \bmath{u}}{t_{\mathrm{stop}}} - \nabla \Phi_{\rm SG} - 2 \eta v_K \Omega \hat{x},
\end{split}
\end{equation}
where $\Phi_{\rm SG}$ is the potential field of the self-gravity of solids. The parameter $\eta$ quantifies the reduction of the gas orbital velocity below the Keplerian value due to radial pressure support in the disk. The final term in the expression containing $\eta$ is a constant radial force in {\sc Athena} to implement the effective global radial pressure gradient under the restrictions of the local model \citep{Bai_Stone_2010a}. For a gas pressure profile $P(R)$, it is given by
\begin{equation}
\eta \equiv -\frac{1}{2}\frac{c_s^2}{v_K^2}\frac{d \ln P}{d \ln R}.
\end{equation}
The self gravity term is obtained by solving Poisson's equation
\begin{equation}
    \nabla^2\Phi_{\rm SG} = 4\pi G \rho_p
\end{equation}
where $G$ is the gravitational constant. We solve this equation using the methods outlined in \cite{Simon_2016} and \cite{Koyama_2009}; in short, we use a Fast Fourier transform in the $x$ and $y$ directions and a Green's function approach for the open vertical boundaries. 

The boundary conditions are shearing-periodic, purely periodic, and modified outflow in the radial, azimuthal, and vertical directions, respectively, the latter of which is setup such that the gas density is extrapolated exponentially into the ghost zones \citep{2010ApJS..189..142S,li_shear}. The outflow boundary condition in the vertical direction results in some mass loss at each timestep  \citep{li_shear}. Therefore at each time step, the gas density is renormalized by multiplying the density in all cells by the mass ratio between the previous and current time steps, so that total gas mass is conserved. During the first $0.5\ \Omega^{-1}$ of simulation, this loss averages to a fraction of $\sim 2\times 10^{-8}$ of the total gas mass in the box; thus on the timescale of a time step (which is less than $0.5\ \Omega^{-1}$), we expect the effect of this mass loss and renormalization to be negligible.

Four dimensionless quantities characterize the physical behavior of the simulations through SI and gravitational collapse: the dimensionless stopping time,
\begin{equation}
    \label{eq:tau}
    \tau = t_{\rm stop} \Omega,
\end{equation}
which relates the particle's aerodynamic and orbital timescales; the global radial pressure gradient parameter,
\begin{equation}
    \label{eq:Pi}
    \Pi = \frac{\eta v_K}{c_s},
\end{equation}
the ratio of solid surface density $\Sigma_p$ to gas surface density $\Sigma_g$,
\begin{equation}
    \label{eq:Z}
    Z = \frac{\Sigma_p}{\Sigma_g},
\end{equation}

\noindent
and the relative strength of self-gravity to tidal shear, which is important during the gravitational collapse of particle clumps,
\begin{equation}
    \label{eq:G}
    \tilde{G} = \frac{4\pi G \rho_0}{\Omega^2},
\end{equation}
where $\rho_0$ is the midplane gas density at $t=0$. This parameter is directly related to the Toomre parameter for a gas disk; $Q = \sqrt{8/\pi}\tilde{G}$.

\subsection{Quantifying Planetesimal Masses} \label{sec:plan_ct}

To identify gravitationally bound clumps inside our domain, we employ the use of the PLanetesimal ANalyzer code \citep[PLAN;][]{Li_2019}. PLAN uses the same approach as HOP \citep{1998ApJ...498..137E}, a code that identifies dark matter halos in cosmological simulations, but applied to pebble clouds instead. First, simulation particles are stored in a data structure based on a Barnes-Hut tree \citep{1986Natur.324..446B}. This allows for fast searches of a given particle's neighbors. Then, particles in dense regions are identified using the HOP algorithm, which is capable of grouping together physically related particles. These particles are grouped with those whose immediate surroundings have a particle density greater than a customized threshold of
$\delta_{\rm outer} = (8/9)\rho_r$ where the Roche density $\rho_r$ is defined to be 
\begin{equation}
    \frac{\rho_r}{\rho_0} = \frac{9}{\tilde{G}}.
\end{equation}
Full details of the algorithm and code are in \cite{Li_2019}.


As discussed in previous works, planetesimals will not collapse further than approximately the grid scale \citep[see][]{Simon_2016}. For the fiducial resolution (see below), a grid cell is $\sim 4 \times 10^{-4}H$, where $H$ is the vertical gas scale height, $H = c_s/\Omega$; at the CCKB radius of 45 AU, this scale corresponds to a grid cell size of approximately $10^5$ km, orders of magnitude larger than even the largest KBOs. Thus, clumps in our simulations will have an enhanced size compared to the actual scales of planetesimals, and we do not expect our simulations to correctly capture the post-formation evolution of planetesimals (e.g., mergers between planetesimals will not, in general, be accurately treated). 

Therefore, we follow \cite{Gole_2020} and employ a clump tracking approach ({\sc ClumpTracker}) to counter these effects. In short, the algorithm tracks clumps from one frame to the next and creates a mass history for each clump. From this data, we can then look at every clump that ever existed in the simulation and measure its mass at the time of formation. Ultimately, measuring initial masses with this method is able to solve three problems stemming from the artificially large size of simulated clumps:

\begin{itemize}
\item Accretion: After formation, clumps will likely accrete particles at an artificially enhanced rate due to the enhanced collisional cross section. Examining only the initial masses ignores this accretion.
\item Mergers: Similarly, clumps will merge at an enhanced rate. If two clumps merge, a snapshot would count them as one large clump. With this tracking method, we measure the initial formation mass of the two separate clumps.
\item Splits and Tidal Stripping: If clumps are able to fully collapse before interacting with other clumps, the rate of clumps being tidally stripped and split into multiple clumps should be quite low. However, this occurs quite often with the artificially large clumps in these simulations, which are less compact and thus more easily tidally stripped. The algorithm is able to ignore this effect by measuring only the initial mass of the parent clump.
\end{itemize}

Note that this is a separate approach than employing sink particles as in e.g., \citet{Schafer2024}. However, \citet{Schafer2024} showed using a clump tracker produces approximately the same IMFs as produced in simulations using sink particles. 

One caveat of the clump tracking algorithm (and the sink particle algorithm) is that it is unable to capture the sub-grid collapse dynamics of the pebble clumps. It is known already that clumps formed via the SI are most likely to form binary (or smaller many-body) objects \citep{2021PSJ.....2...27N}. We are unable to resolve this collapse, and there is no known method of calculating the final number of bodies from clump properties alone. As a workaround, we implement a toy model to approximately model this sub-grid collapse using the results from \cite{2021PSJ.....2...27N} (described in Section~\ref{sec:binary}). We find that it does not lead to drastic changes in the size distribution. Therefore, for the purposes of this paper, we assume that each pebble cloud collapses into one single planetesimal; we refer to this as monolithic collapse.

\subsection{Parameters, Initial Conditions, and Numerical Setup} 

In this work, we isolate the impact of each physical parameter ($\tau$, $Z$, $\Pi$, $\tilde{G}$) by varying them individually while holding the others fixed. We chose these parameters because of their identified significance in previous IMF studies \citep{Simon_2017,Abod_2019,Li_2019}.  In addition to physical parameters, several numerical factors are known to influence the resulting IMF, such as resolution and domain size \citep{Simon_2016, 2021MNRAS.500..520R}.  The fiducial simulation uses a domain size slightly larger than in previous studies \citep{Simon_2016, Simon_2017, Abod_2019}, with dimensions $L_x \times L_y \times L_z = 0.4H \times 0.2H \times 0.2H$. The corresponding fiducial resolution is $N_x \times N_y \times N_z = 1024 \times 512 \times 512$, and the total number of particles is fixed at $N_{\rm tot} = 2.7 \times 10^8$. At initialization, this setup results in an average of one particle per grid cell. As the simulation progresses, particles settle toward the midplane due to vertical gravity of the star, leading to an increased number of particles per cell around the midplane.

In addition to simulation size/scale, the simulation duration and timing have the potential to affect the IMF. Each simulation was run to $t_{\rm end}$, with slight variations in the self-gravity activation time $t_{\rm SG}$. The extent to which differences in $t_{\rm end}$ affect resulting IMFs will be addressed in Section~\ref{sec:unc_len}. 

Importantly, while we decrease the resolution for two of our simulations to test the effect of changing the grid scale, $N_{\rm tot}$ is kept constant. Therefore, decreasing the resolution by a factor of two in each spatial direction results in an average of eight particles per grid cell at initialization. Previous studies have shown that, provided sufficient resolution is achieved, the onset and nonlinear evolution of the SI are not strongly sensitive to particle resolution \citep{2007ApJ...662..627J, Bai_Stone_2010b, 2017A&A...606A..80Y}. A summary of the parameters used in each simulation is provided in Table~\ref{tab:run_list}, where the naming convention indicates the parameter combinations used in the simulation (e.g. \_Pi05\_ means $\Pi=0.05$, and \_Z5\_ means $Z=0.05$, etc.). 

The particles are initially distributed uniformly in $x$ and $y$ and with a Gaussian profile in $z$. They are initialized with NSH velocities \citep{1986Icar...67..375N} in $x$ and $y$, and vertical velocities are initialized to zero. While we add random noise to the particle locations to seed the SI, it is worth mentioning that the particles are not initially in an equilibrium state in $z$, and experience a transient settling phase early in the simulation.

\begin{deluxetable*}{lcccccccc}[htb!]
    \label{tab:run_list}
    \tablecaption{List of simulations}
    \tablehead{ \colhead{Name} & \colhead{$N_x \times N_y \times N_z$} & \colhead{$\tau$} & \colhead{$\Pi$} & \colhead{$Z$} & \colhead{$\tilde{G}$} & \colhead{$t_{\rm SG}$ ($\Omega^{-1}$)} & \colhead{$t_{\rm end}$ ($\Omega^{-1}$)}}
    \startdata
         \Fiducial\tablenotemark{a} & $1024\times512\times512$  & 0.1 & 0.05 & 0.05 & 0.1 & 30 & 40  \\
         \LowR & $512\times256\times256$  & 0.1 & 0.05 & 0.05 & 0.1 & 30 & 40  \\
         \UltraLowR & $256\times128\times128$  & 0.1 & 0.05 & 0.05 & 0.1 & 30 & 40  \\
         \LowZ & $1024\times512\times512$  & 0.1 & 0.05 & 0.02 & 0.1 & 30 & 58  \\
         \LowG & $1024\times512\times512$  & 0.1 & 0.05 & 0.05 & 0.05 & 30 & 40  \\
         \HighP & $1024\times512\times512$ & 0.1 & 0.1 & 0.05 & 0.1 & 30 & 45  \\
         \LowT & $1024\times512\times512$  & 0.01 & 0.05 & 0.05 & 0.1 & 200 & 296  \\
         \Gole & $1024\times512\times512$ & 0.3 & 0.05 & 0.02 & 0.1 & 20 & 40  \\
    \enddata

    \tablecomments{The name and parameters that were varied are listed for each simulation. The parameters are dimensionless stopping time $\tau$ (Equation~\ref{eq:tau}), pressure gradient $\Pi$ (Equation~\ref{eq:Pi}), dust-to-gas ratio $Z$ (Equation~\ref{eq:Z}), strength of self-gravity $\tilde{G}$ (Equation~\ref{eq:G}), and the times at which  self gravity is activated $t_{\rm SG}$ and the simulation ends $t_{\rm end}$. The dimensions of each shearing box are $0.4H \times 0.2 H \times 0.2 H$. There are several simulations whose $t_{\rm SG}$ and $t_{\rm end}$ values differ from \Fiducial.  They are \LowZ, which had to run longer in order to reach a state in which the number of emerging planetesimals no longer materially affects the resulting IMF (see in Section~\ref{sec:unc_len}), \LowT\ which took much longer for SI clumping to begin before SG activation, in addition to requiring a longer run time for the same reasons as \LowT\,and \Gole, which was designed to match the parameters of the fiducial run in \cite{Gole_2020}. 
    \tablenotetext{a}{denotes that the \Fiducial\ runs were performed 3 individual times with different random particle seeding, in order to assess stochasticity (see Section~\ref{sec:unc_sto} for more details). }
    }

\end{deluxetable*}

\subsection{Disk Model} \label{sec:disk_model}

In order to establish a connection between our simulations, which are in code units, and the actual physical units of CCKBOs, we employ a disk model. This model establishes $\rho_g$, $H$, $\Omega$, and $c_s$ as functions of position within the disk, the mass of the central star, and the orbital radius at which we place our shearing box domain.  We chose an exponentially tapered surface gas density profile \citep{10.1093/mnras/168.3.603},
\begin{equation} \label{eq:profile}
    \Sigma_g = \Sigma_c \left( \frac{R}{R_c}\right)^{-\gamma} \exp \left[-\left(\frac{R}{R_c}\right)^{2-\gamma}\right];
\end{equation}
$\gamma$ is the power law index, $R_c$ is the disk's characteristic radius at which the exponential tapering begins to dominate, and
\begin{equation}
    \Sigma_c = (2-\gamma) \frac{M_d}{2\pi R_c^2},
\end{equation}
where $M_d$ is the total gas mass of the disk. 


Observations of circumstellar disks indicate that  $\gamma$ is close to 1 \citep{2009ApJ...700.1502A,2010ApJ...723.1241A}, though a  range between $0.4 - 1.1$ is observed across systems, and the characteristic radius $R_c$ typically lies between $\sim$10 and 100 AU \citep{2010ApJ...723.1241A}. Models of the solar nebula often explore a broad parameter space rather than a single fiducial profile. In particular, investigations of models for  the parameters of the solar nebula fit $\gamma \sim 1$, $R_c \leq$ 50 AU, and disk masses $M_d \leq 0.1M_\odot$, while still allowing for wider ranges to reflect observational and theoretical uncertainties \citep{2012AJ....143...91K, solar_nebula}. 

Assuming a vertically Gaussian density profile (which itself assumes a temperature that depends only on the cylindrical radius from the star; see below), we define the gas density in terms of $\Sigma_g$:
\begin{equation}
    \rho_g = \frac{\Sigma_g}{\sqrt{2\pi}H},
\end{equation}
We also convert between physical and code mass units by the mass:
\begin{equation}
\label{eq:mh}
    M_H = \rho_gH^3,
\end{equation}
using
\begin{equation} \label{eq:omega}
    \Omega = \sqrt{\frac{GM_\odot}{R^3}}
\end{equation}
and
\begin{equation} \label{eq:soundspeed}
    c_s = \sqrt{\frac{k_BT}{\mu m_H}},
\end{equation}
where $k_B$ is the Boltzmann constant, $T$ is temperature, $\mu$ is the mean molecular weight of 2.37, and $m_H$ is the mass of hydrogen.

We adopt a radial temperature profile for the disk based on the formulation presented in \cite{2020ALMA}, while neglecting a vertical dependence:
\begin{equation} \label{eq:temperature}
    T = 19\ {\rm K} \left(\frac{R}{{30\ \rm AU}}\right)^{-q}.
\end{equation}
The constants are chosen such that the CO snowline, corresponding to a temperature of 19 K, and lies at Neptune’s orbital distance of 30 AU \citep{COSnowLine}. The exponent $q$ typically takes a value around $0.5$ from passively irradiated disks \citep{2001ApJ...560..957D}, though it can vary depending on disk conditions \citep[][treats it as a free parameter]{2020ALMA}, and some extrasolar protoplanetary disks suggest $q \gtrsim 0.5$ \citep{2025ApJ...984L..10G}. 

With this formulation, $M_H$ is now expressed as a function of the disk model parameters. Some choices are straightforward—for example, setting $\gamma = 1$ and choosing an orbital radius of $R=45$ AU, matching the CCKB. This reduces our $M_H$ to a function of three disk model parameters: 

\begin{equation} \label{eq:MH}
\begin{split}  
M_H = 2\times10^{-5}\,
\left(\frac{M_d}{0.1 M_\odot}\right)\,
\left(\frac{R_c}{45\,{\rm AU}}\right)^{-1}
\left(\frac{3}{2}\right)^{-q}\\
\times \exp\left(-\frac{45\,{\rm AU}}{R_c}\right)\ M_\odot,
\end{split}
\end{equation}


This disk model is crucial to comparing observations and simulations when parameterizing distributions, and its uses and uncertainties are explored further in Section~\ref{sec:unc_disk}.

\subsection{Model Fitting} \label{sec:model}

We fit parametrized models to the planetesimal IMF using a Markov Chain Monte Carlo (MCMC) approach implemented with the \texttt{emcee} package \citep{mcmc}. Unlike approaches based on maximum-likelihood estimation followed by bootstrap resampling, such as \citet{Li_2019}, MCMC directly samples the posterior probability distributions of the model parameters, from which both parameter estimates and uncertainties can be obtained.


We assume that the observed KBO masses, $\bmath{M}={M_1,\ldots,M_N}$, are independent draws from a probability density function $p(M\mid\bmath{\theta})$, where $\bmath{\theta}$ is the vector of model parameters. The likelihood is therefore:
\begin{equation}
\mathcal{L}(\bmath{M}\mid\bmath{\theta}) \equiv
p(\bmath{M}\mid\bmath{\theta}) = \prod_{i=1}^{N} p(M_i\mid\bmath{\theta}).
\end{equation}
Bayes' theorem gives the posterior:
\begin{equation}
p(\bmath{\theta}\mid\bmath{M})
\propto
p(\bmath{M}\mid\bmath{\theta})p(\bmath{\theta}),
\end{equation}
where $p(\bmath{\theta})$ is the prior. The prior encodes our knowledge about the distribution before drawing from it. Note that the proportionality is sufficient for parameter estimation; the MCMC algorithm determines $\bmath{\theta}$ values from the posterior $p(\bmath{\theta}\mid\bmath{M})$ in a way that is independent of the normalizing constant.

In practice, it is easier to work with a logarithmic mass coordinate, $x = \ln (M/M_{\rm min})$, where $M_{\rm min}$ is the minimum mass of the distribution. We find it convenient to express the PDF as
\begin{equation}
    p(x\mid\bmath{\theta}) = c(\theta)f(x\mid\bmath{\theta})
\end{equation}
where $c(\bmath{\theta})$ is a normalization factor. The CDF is then
\begin{equation}
\begin{split}
    P_>(x\mid\bmath{\theta}) &= \int_x^{+\infty} p(x'\mid\bmath{\theta})dx' \\
    &= c(\bmath{\theta}) \int_x^{+\infty} f(x'\mid\bmath{\theta})dx',
\end{split}
\end{equation}
which shows the expected fraction of clumps with masses larger than $M = e^xM_{\rm min}$. The normalization factor is chosen such that

\begin{equation}
    \int_0^{+\infty} p(M\mid\bmath{\theta})dM =
    \int_{-\infty}^{+\infty} p(x'\mid\bmath{\theta})dx' = 1
\end{equation}

\noindent
In other words,

\begin{equation}
    c(\bmath{\theta}) = \frac{1}{\int_{-\infty}^{+\infty}f(x\mid\bmath{\theta})dx},
\end{equation}
and requires that $f$ does not diverge as $x \rightarrow +\infty$.

Putting these together, the log-likelihood is then
\begin{equation}
\begin{split}
    \ln \mathcal{L}(\bmath{x}\mid\bmath{\theta}) &= \sum_{i=1}^{N} \ln p(x_i\mid\bmath{\theta})\\ 
    &= N \ln c(\bmath{\theta}) + \sum_{i=1}^{N} \ln f(x_i\mid\bmath{\theta}).
\end{split}
\end{equation}

The posterior distributions {can be characterized with a central estimate and credible interval. In order to retain the often skewed shapes of these distributions, this study utilizes the median and the 0.16 and 0.84 quantiles as the uncertainty. This is an improvement over the common practice of reporting a standard deviation, which suggests a symmetry that often isn't present. Moreover, medians and quantiles are \textit{robust statistics}, meaning that they are robust to outliers (unlike the commonly used means and standard deviation). Note also that some distributions are truncated at the parameter boundary by the priors imposed. For these cases, we represent them in the same way (median plus quantiles) and comment on how our priors influenced the posterior distribution.

Typical MCMC chains consist of 24 walkers and 11,000 steps. We discard the first 1,000 steps as burn-in and assess convergence using the integrated autocorrelation time, $\tau$, of each model parameter. For all fits, the retained chains extend for at least $N_{\rm steps}/\tau_{\rm max} > 100$ autocorrelation times, where $\tau_{\rm max}$ is the largest autocorrelation time among the fitted parameters. This corresponds to an effective sample size of the MCMC chain ($N_{\rm eff}$) of approximately $N_{\rm eff} \gtrsim 1200$. As an additional stability check, posterior medians vary by less than 1\% when calculated from progressively longer portions of the retained chains. 

Unlike \cite{Li_2019} and \cite{Gole_2020}, we will focus on optimizing one model, the \textit{Variably Tapered Power Law} \citep[VTPL,][]{2017A&A...597A..69S}. Although \citet{Gole_2020} favored models other than the VTPL, we adopt the VTPL because it reflects the exponentially tapered size distribution observed in the CCKB, as inferred from the debiased Outer Solar System Origins Survey ensemble sample \citep[OSSOS or OSSOS++]{Kavelaars}, and the DECam Ecliptic Exploration Project \citep[DEEP]{Napier}. More recent observational studies favor different power laws with fewer parameters for the smaller-sized CCKBOs \citep{Bernstein2026, Eduardo2026}, but since the numerical resolution of the SI simulations will only create the largest CCKBOs, we adhere to the tapered power law prescriptions, which has more descriptive power for the larger objects.

The VTPL is expressed as a cumulative probability distribution $P(>M)$ that can be converted to a cumulative histogram $N(>M)$ by multiplying by the total number of planetesimals. The VTPL is expressed in terms of 3 parameters, $\bmath{\theta} = (\alpha_M, \beta_M, M_{\rm exp})$, such that,
\begin{equation} \label{eq:VTPL}
    P(>M, \bmath{\theta}) = c M^{-\alpha_M} \exp \left[ - \left( \frac{M}{M_{\rm exp}}\right)^{\beta_M}\right],
\end{equation}
and the normalization coefficient $c$, chosen such that $P(>M_{\rm min}) =1$, is
\begin{equation}
    c = M_{\rm min}^{\alpha_M} \exp \left[ \left(\frac{M_{\rm min}}{M_{\rm exp}}\right)^{\beta_M} \right].
\end{equation}

The corresponding probability density is the negative derivative of the cumulative distribution,
\begin{equation}
p(M\mid\bmath{\theta})
= -\frac{dP(>M\mid\bmath{\theta})}{dM},
\end{equation}
which gives
\begin{equation}
\begin{split}
p(M\mid\bmath{\theta}) =
cM^{-(\alpha_M+1)}
& \exp\left[
-\left(\frac{M}{M_{\rm exp}}\right)^{\beta_M}
\right]
\\ &\times \left[
\alpha_M+
\beta_M\left(\frac{M}{M_{\rm exp}}\right)^{\beta_M}
\right].
\end{split}
\end{equation}
The resulting log-likelihood is therefore
\begin{equation}
\begin{split}
\ln \mathcal{L}
= \sum_i^{N_{\rm plt}} \Bigg[
&\ln\left(
\alpha_M+
\beta_M\left(\frac{M_i}{M_{\rm exp}}\right)^{\beta_M}
\right) \\
&-\left(\frac{M_i}{M_{\rm exp}}\right)^{\beta_M}
-(\alpha_M+1)\ln M_i
\Bigg]
+ N_{\rm plt}\ln{c}
\end{split}
\end{equation}

We allow either $\alpha_M$ or $\beta_M$ to be negative, \dc{but} impose two conditions\dc{:} First, at least one of $\alpha_M$ or $\beta_M$ must be positive so that $P(>M)\rightarrow0$ as $M\rightarrow\infty$. Second, the PDF must remain non-negative over the fitted mass range. Since all other factors in $p(M\mid\bmath{\theta})$ are positive, this requires
\begin{equation}
\alpha_M+
\beta_M\left(\frac{M}{M_{\rm exp}}\right)^{\beta_M}>0.
\end{equation}
The quantity on the left-hand side increases monotonically with $M$ (even for $\beta_M < 0$), so it is sufficient to impose this condition at the lower bound,
\begin{equation}
\alpha_M+
\beta_M\left(\frac{M_{\rm min}}{M_{\rm exp}}\right)^{\beta_M}>0.
\end{equation}
This constraint ensures that the VTPL cumulative distribution decreases monotonically with increasing $M$ and that its corresponding PDF is physically valid throughout the fitted mass range.



The priors also include bounding \M\ between $M_{\rm min} $ and $M_{\rm max}$. Since the exponential taper mass $M_{\rm exp}$ is strictly positive and may span several orders of magnitude, we adopt a log-normal prior. Specifically, we assume that $\ln M_{\rm exp}$ follows a normal distribution,
\begin{equation}
    \ln M_{\rm exp} \sim \mathcal{N}(\bar{M}, \sigma^2),
\end{equation}
We adopt $\sigma = 1$, corresponding to a weakly informative prior that permits $\ln M_{\rm exp}$ to vary by a factor of $\sim e$ about its central value in one standard deviation. This choice is not intended to impose a strong constraint on the parameter, but rather to provide a broad prior that regularizes the inference while allowing the likelihood to dominate.  

The log of the prior is 
\begin{equation}
    \ln p(\bmath{\theta}) = -\frac{(\ln M_{\rm exp} - \bar{\mu})^2}{2\sigma^2} - \ln(M_{\rm exp}\,\sigma\sqrt{2\pi}),
\end{equation}
where $\bar{\mu}$ sets the central value of the prior in log space. We adopt $\bar{\mu} = \ln(0.8) - 0.5$, which places the mean of the log-normal prior at 0.8, while remaining sufficiently broad that it does not significantly influence the posterior. We also attempted to use a uniform distribution prior between $M_{\rm min}$ and $M_{\rm max}$, instead of the log-normal. The uniform prior led to near identical posterior distributions on \M, but the sampler would converge in significantly more steps, therefore we only use the log-normal for fitting.

\subsection{Observational Overview and Constraints}

We highlight comparisons between 4 observational studies containing 7 fits to CCKBO observational magnitude distributions \citep{Kavelaars, Napier, Eduardo2026, Bernstein2026}. For convenience in text and figures, we abbreviate the fits to the author's first initial and last two digits of the publication year with a hyphen to differentiate multiple fits in a single study. The selected observational fits are as follows:
\begin{enumerate}
    \item \citet{Kavelaars}: Outer Solar System Origins Survey (OSSOS++) survey containing CCKBOs between $5.0 \lesssim H_r \lesssim 8.3$, debiased and fit with a tapered power law assuming two different values of $\alpha_{H_r}$: $\alpha_{H_r} = 0.4$ (K21-1), and $\alpha_{H_r} = 0.5$ (K21-2).
    \item \citet{Napier}: DECam Ecliptic Exploration Project (DEEP) survey containing CCKBOs with $H_r \leq 10.5$ (N24).
    \item \citet{Eduardo2026}: CCKBOs observed by NIRCam aboard the JWST containing objects $8.3 \lesssim H_r \lesssim 14.0$, fit to two different samples: one with all three epoch detections (sample A, E26-1 and another based on the combined set of three-epoch and two-epoch detections (sample A+B, E26-2).
    \item \citet{Bernstein2026}: Combined survey from Minor Planet Center (MPC), OSSOS, HST, and JWST spanning $5.0 \lesssim H_r \lesssim 13.0$, where MPC objects are selected with two different limits to create two tapered power law fits: \textit{bright-app} (B26-1) and \textit{bright-abs} (B26-2).
\end{enumerate}

To facilitate a direct comparison between the observed magnitude distributions and the planetesimal mass outputs from numerical simulations, both magnitude and mass can be converted to diameter to compare size distributions. These considerations warrant several precise definitions. For the purposes of this study, an IMF is defined as the distribution of mass values from a simulation, a predicted size distribution (PSD) is the size distribution (in diameter) of the simulated clumps undergoing monolithic collapse, and the observational distribution refers to the diameter sizes of objects within a debiased $H_r$ magnitude distribution of the CCKB. These definitions require adopting consistent assumptions regarding the properties of the CCKB population.

Observed CCKB absolute magnitudes in the $r$-band ($H_r$) can be converted to diameters (in kilometers) using the following relation:
\begin{equation} \label{eq:Htor}
    D_{\rm km}(H_r)  = 
    \frac{1329}{\sqrt{{\rm albedo}}}10^{0.2*H_r},
\end{equation}
where an assumed geometric albedo is required, in general, 0.15.

Similarly, to convert planetesimal mass $M$ to diameter (in kilometers), we use the aforementioned monolithic collapse assumption, where the mass collapses into a single spherical body with a given bulk material density $\rho_{\rm mat} = 1\ {\rm g}/{\rm cm}^3$ (this choice is discussed in Section~\ref{sec:obs_unc}). The corresponding diameter is given by:
\begin{equation} \label{eq:rtom}
    D_{\rm km}(M) = \left[\frac{6}{\pi\rho_{\rm mat}} M\right]^{1/3}.
\end{equation}

\section{Results} \label{sec:results}

\subsection{Primary Result} \label{sec:primary}

Since both \citet{Kavelaars} and \citet{Napier} identified the unique exponential tapered shape of the distribution of CCKBO magnitudes/size, one of our primary goals is to examine if the SI is capable of reproducing this taper.  Here we apply normalization techniques to isolate the shape of the PSD; this eliminates the use of a disk model and the assumption of a number of parameters, such as albedo and material density of planetesimals. Subsequent comparisons will reintroduce the physical scaling in order to test the simulations with fitted distribution parameters.

The most widely used normalization is the z-score transformation, which standardizes a distribution by expressing each value in terms of the number of standard deviations it lies from the mean. This is given by:
\begin{equation}
    z_i = \frac{x_i - \bar{x}}{\sigma}
\end{equation}
where $z_i$ is the standardized value for the $i$-th observation, $x_i$ is the original value, $\bar{x}$ is the mean of the distribution, and $\sigma$ is the standard deviation.

Because the z-score transformation is a linear transformation, it preserves the relative spacing and ordering of the data. As a result, the overall shape of the distribution -- such as its symmetry, skewness, and modality -- remains unchanged. Further details of this normalization will be discussed in subsequent sections.

In Figure~\ref{fig:first}, the z-scored PSDs from each simulation are shown in color, overlaid on samples of an observed CCKB distribution (specifically \citealt{Kavelaars}, in black), which are also z-scored.  A full description of the sampling process and the justification for its use is explained in Section~\ref{sec:obs}. The observation sample and PSDs are each normalized by the total number of sampled objects. The error bars on the CCKB distribution represent the expected uncertainties resulting from \textit{only} the debiasing procedure described in \citet{Kavelaars}. 

\begin{figure*}[htb!]
    \centering
    \includegraphics[width=1\linewidth]{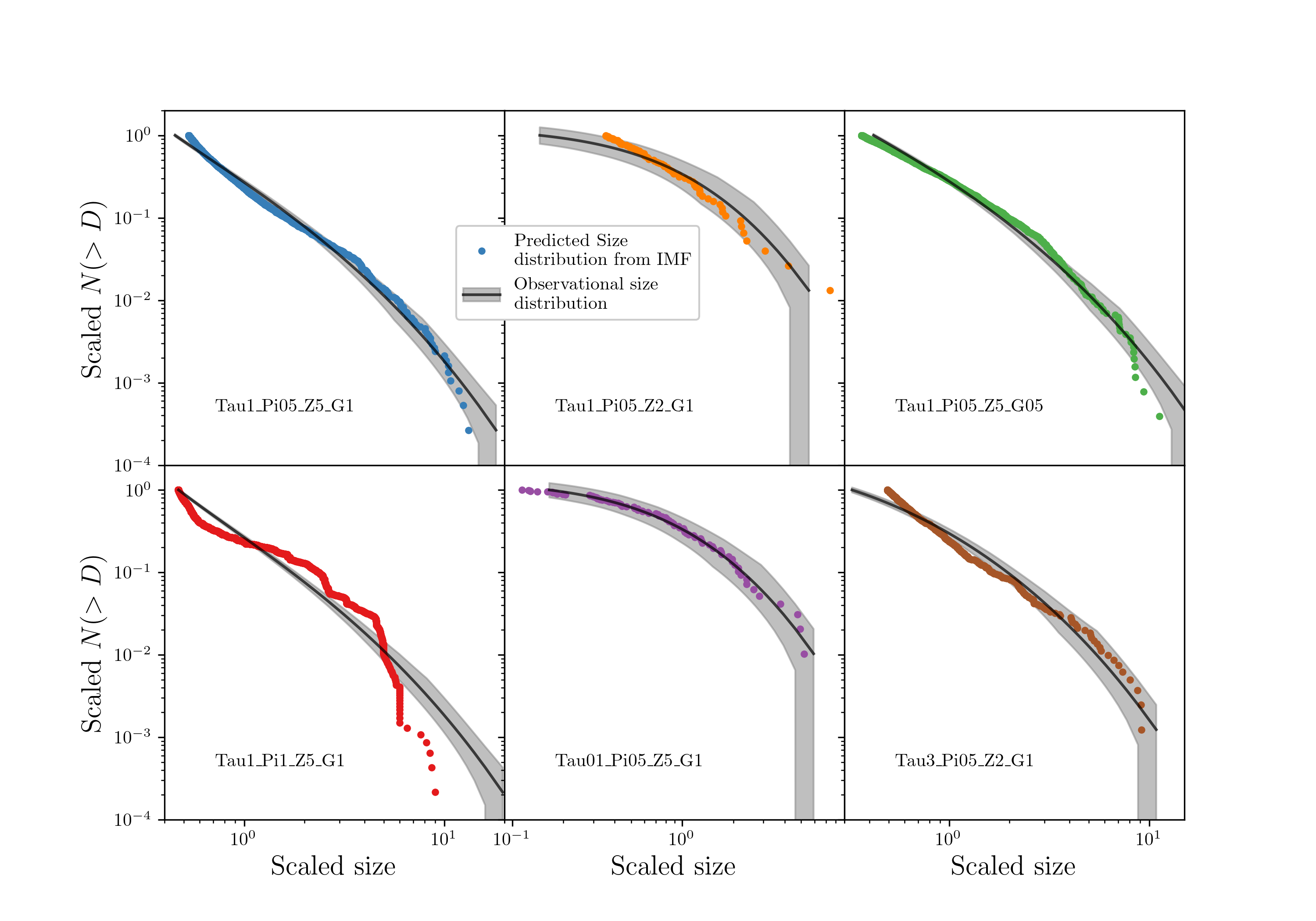}
    \caption{The normalized predicted size distributions for each simulation (color dots) and a representative sample of the $\alpha = 0.66$ fit to the CCKB from \citet{Kavelaars} (black lines) systematically sampled to match the number of objects in the PSD within its range, with 95\% Poisson confidence from debiasing (grey shaded regions). The PSD and CCKB sample are normalized via the z-score, a linear transformation that converts the values in a distribution to the number of standard deviations away from the mean, then arbitrarily shifted by 1 to be positive on the $x$-axis.  All $N(>D)$ have been scaled down by the total number of objects in the respective panel for visual clarity. Note the changing $x$-axis limits for each panel. There is very strong agreement between predictions and observations for \LowT~(lower middle), \LowG~(upper right), \Fiducial~(upper left), and to a lesser extent, \LowZ~(upper middle).}
    \label{fig:first}
\end{figure*}

The visual similarities between simulations and observations in Figure~\ref{fig:first} are striking. Namely, the \LowT size distribution is completely encompassed by observational debiasing uncertainty. Multiple simulation size distributions are largely encompassed by the observational debiasing uncertainty: \LowZ, \LowG, and two more are just outside the uncertainty edges: \Fiducial, \Gole. We will now quantify this agreement. 


\subsection{Statistical Analyses} \label{sec:obs}

The z-score removes the need for assumptions regarding a disk model; however, an additional challenge is the issue of completeness. Ideally, the simulations would generate a sufficient number of planetesimals to represent the complete population expected in a primordial CCKB (i.e. the total numbers match). In practice, however, it is not feasible for a single shearing-box simulation to achieve this, as it represents only a small patch of the disk for only a brief period in time. In addition, the smallest objects formed in the simulations, which are roughly the scale of $\Delta x$, are likely underrepresented. As a result, we remove them for a more robust comparison (see Appendix~\ref{sec:resolution_effects} for full discussion). Since the number of objects in the PSD is less than the predicted number of objects at the same size in observation, it becomes necessary to sample the observational fits to match PSD numbers.

Statistical analysis will be used to confirm the visual agreement in Figure~\ref{fig:first} in Section~\ref{sec:first_comp}, and finally a statistical comparison with the parametrized model described is presented in Section~\ref{sec:vtpl}.

\subsubsection{Kolmogorov-Smirnov Test} \label{sec:first_comp}

A good tool to characterize the relationship between distributions and obtain a single metric is with a Kolmogorov-Smirnov (K-S) test. The one-sample K-S test takes a given sample and a given distribution and calculates the probability (see below for a more detailed explanation of what this probability represents for K-S tests) that said sample was drawn from that exact distribution. The two-sample K-S test calculates the probability that two samples come from the same underlying distribution, without saying anything about what that distribution is. For both one-sample and two-sample, the test statistic ($D_{KS}$) for the K-S test is the largest absolute difference between the two distributions across all values, and a $p$-value is calculated by comparing this distance to the Kolmogorov distribution. Since samples can be normalized, we will use the two-sample K-S test, which requires constructing a sample from the analytical fit to the observational distributions. 


To construct a sample from their analytical distribution, \citet{2023ApJ...949...81G} samples $N_{\rm max}$ objects from the IMF, where $N_{\rm max}$ is the total number of objects in the simulation IMF, between $M_{\rm min}$ and $M_{\rm max}$. This is repeated to create a new $N_{\rm max}$ sample many times, then each object's mass is averaged across its ordinal position in the sorted sample to create a representative sample of the analytical distribution. We reuse this sampling process for the CCKB distribution in Figure~\ref{fig:first} (with $D_{\rm min}$ and $D_{\rm max}$ from the PSD instead of $M_{\rm min}$ and $M_{\rm max}$ from the IMF), which we referred to as a representative sample. After the representative sample is constructed, it and the PSD sample are normalized via z-score and evaluated with the two-sample K-S test.
Table~\ref{tab:kstest} summarizes the results of two-sample K-S tests (via \verb|scipy.stats|) between the PSDs and the 7 fits to observations \citep{Kavelaars, Napier, Eduardo2026,Bernstein2026}.

\begin{deluxetable*}{c c c c c c c c c}[htb!]
\tablecaption{two-sample K-S Test results}
\tablehead{
 & \colhead{\# of Objects} & \colhead{K21-1} & \colhead{K21-2} & \colhead{N24} & \colhead{E26-1} & \colhead{E26-2} & \colhead{B26-1} & \colhead{B26-2}
}
\startdata
\Fiducial & 3762 & $\sim0$ & $\sim0$ & $\sim0$ & $\sim0$ & $\sim0$ & $\sim0$ & $\sim0$\\
\LowZ & 76 & 0.003 & $\sim0$ & 0.003 & 0.006 & 0.006 & $\sim0$ & $\sim0$\\
\LowG & 2567 & $\sim0$ & {\bf 0.198} & 0.035 & $\sim0$ & $\sim0$ & $\sim0$ & $\sim0$\\
\HighP & 4644 & $\sim0$ & $\sim0$ & $\sim0$ & $\sim0$ & $\sim0$ & $\sim0$ & $\sim0$\\
\LowT & 97 & {\bf 0.899} & {\bf 0.801} & {\bf 0.899} & {\bf 0.964} & {\bf 0.993} & {\bf 0.993} & {\bf 0.801}\\
\Gole & 808 & $\sim0$ & $\sim0$ & $\sim0$ & $\sim0$ & $\sim0$ & $\sim0$ & $\sim0$\\
\enddata
    \tablecomments{The two-sample K-S test $p$-values between simulation and corresponding sample of observational population. Those with $p$-values greater than 0.05 are in bold, and those with $p$-values $<10^{-6}$ are listed as $\sim 0$. A $p$-value less than 0.05 means that the K-S test rejects the null hypothesis that the simulated and observed samples were drawn from the same underlying distribution. Note that the small MPC sample lacks the power to reject any of our simulations, but the OSSOS and DEEP surveys, with their much larger data sets, reject every model except for \LowT\ and \LowG\ when compared with K21-2.}
    \label{tab:kstest}
\end{deluxetable*}


We present all of these $p$-values in Table~\ref{tab:kstest}. It is important to note that the K-S test has high specificity (it can reject the null hypothesis $H_0$ that two samples came from the same population) but low sensitivity (failure to reject $H_0$ does not tell you that the samples came from the same population). The large sample sizes such as those of the OSSOS and DEEP surveys have significant discriminating power and solidly reject every model except for \LowT, which retains a high $p$-value across the board. It should be noted that this method considers only the observational distribution described by the mean or median parameters presented in each study. Sampling does not take into consideration uncertainty in parameters due to fitting, or the uncertainty from debiasing the CCKB population. In other words, this test is designed to be rigid or conservative when comparing to observation. Despite this, combined with the appearance of Figure~\ref{fig:first}, it is apparent that the \LowT\ simulation produced a planetesimal PSD that is similar to observational distributions.



\subsubsection{Parametrized Comparison} \label{sec:vtpl}

Previous sections analyzed the comparison between the PSDs and samples from the observational distributions. We will now consider parameterized fits to the PSDs and compare them with the fits to the observed distributions.  The goal is to determine if the shape (i.e. the exponential taper) of the distributions in \citet{Kavelaars}, \citet{Napier}, \citet{Eduardo2026}, and \cite{Bernstein2026} are recreated by the simulation PSDs. 

Section~\ref{sec:model} described the parametrized model fitting technique employed in this section to characterize an IMF. For a PSD, it is instead characterized in size space by $\bmath{\theta} = (\alpha, \beta, D_{\rm exp})$, which is a transformation of the mass space and magnitude ($H_r$) space parameters \citep[those of][]{Kavelaars, Napier}: 
\begin{equation}
    \alpha = 3\alpha_M = 5\alpha_{H_r}
\end{equation}
\begin{equation}
    \beta = 3\beta_M = 5\beta_{H_r}
\end{equation}
\begin{equation}
\label{eq:dexp_hb}
    D_{\rm exp} = \left(\frac{6}{\pi\rho_{\rm mat}}M_{\rm exp}\right)^{1/3} = \frac{1329}{\sqrt{{\rm albedo}}}10^{-0.2H_B}.
\end{equation}


This parameterized model fitting (specifically the value of \D) must rely on disk model parameters ($M_H$), albedo, and CCKBO bulk density, since it is performed in physical unit space, and cannot utilize the z-score normalization. The mapping between PSDs and the observed CCKB size distribution is degenerate, as different combinations of albedo, $\rho_{\rm mat}$, and disk parameters ($M_d$, $R_c$, $q$) can yield similar physical size scales. In this work, we adopt fiducial values of an albedo of 0.15 and a bulk density of $1\ {\rm g\ cm^{-3}}$, and consider disk models spanning plausible solar nebula conditions \citep[c.f.][]{solar_nebula}. 
For a given PSD, we determine the value of $M_H$ required to align the distribution with the observed $N(>D)$, and infer the corresponding disk parameters ($M_d$, $R_c$, $q$; see Table~\ref{tab:model_values} and Appendix~\ref{app:dm}). The disk model is therefore used solely to provide a physical interpretation of the inferred mass scale. For all simulations, we perform this alignment by matching the 29th largest object in the PSD to the smallest object in the MPC sample, since the MPC sample contains 29 objects. It is possible to perform a one-sample K-S test between the aligned PSD and the observational distributions, but the K-S test statistic is strongly tied to the value chosen for $M_H$. Because the absolute mass scaling $M_H$ is uncertain, we do not calculate any one-sample K-S test between PSD and observational distribution.

Upon performing this alignment, a significant issue becomes apparent. In every simulation, the smallest planetesimal, $D_{\rm min}$, is larger than any of the characteristic sizes \D\ inferred from the observations (corresponding to the observational parameter $H_B$; see Equation~\ref{eq:dexp_hb}). Moreover, when a VTPL is fit to a PSD following Section~\ref{sec:model}, the imposed priors truncate the \D\ posterior at $D_{\rm min}$. Thus, for every simulation considered here, essentially the entire resolved planetesimal population lies in the taper-dominated regime of the size distribution, governed by $\beta$, rather than in the power-law regime governed by $\alpha$. Consequently, $\alpha$ is only weakly constrained by fits to the PSDs because very few, if any, simulated planetesimals lie below $D_{\rm exp}$, where the power-law component is most relevant. More importantly, an $\alpha$ inferred from a PSD would characterize a fundamentally different size range from an $\alpha$ inferred from the observations, since $D_{\rm min,sim} > D_{\rm exp,obs}$.

\citet{Bernstein2026} notes a related discrepancy between previous SI studies and their observational fits, finding that the CCKB size distribution favors substantially smaller taper parameters ($\beta$) than those inferred from earlier SI simulations. The limited size range sampled by the simulations may help explain this difference. Because the simulated planetesimals lie predominantly in the taper-dominated regime, a full VTPL fit can yield a poorly constrained and artificially steep $\alpha$. Since $\alpha$ and $\beta$ jointly determine the shape of the VTPL, bias in the inferred $\alpha$ can in turn shift the fitted value of $\beta$, complicating previous direct comparisons between simulation and observational fits.

This suggests that there may be more potential in allowing \D\ to be outside the bounds of the PSD, and constraining it, along with $\alpha$, to observational values. This allows us to focus on $\beta$, which dominates the large-mass/size edge of the distribution and is better probed by our simulations. We can also use this as an exercise to see if the observationally inferred parameters create a better fit to the PSDs. Therefore, we constrain $\alpha$ and \D\ to the seven combinations of values determined by fits to observational data, performing seven fits to $\beta$ alone for each PSD. These fits are presented in Tables~\ref{tab:constrained_Hill_fits} and \ref{tab:constrained_Hill_fits_cont}, and the parameter comparison is visualized in Figure~\ref{fig:constrained_Hill_fits}. 

To test whether the VTPL where we only vary $\beta$ is a reasonable fit, we can use the one-sample K-S test between the PSDs and the one-parameter VTPL, and interpret the K-S test statistic $D_{\rm KS}$, representing the maximum absolute separation between the empirical and model cumulative distribution functions, to quantify the distributional discrepancy. In general, fewer than 5 percentage points in discrepancy ($D_{\rm KS} < 0.05$) is considered a ``good fit''. The values for each PSD and fit are presented in parentheses for in Table~\ref{tab:constrained_Hill_fits}.

\begin{deluxetable*}{ccccc}[htb!]
\tablecaption{VTPL fits to $\beta$ alone}
\tablehead{ & \colhead{K21-1} & \colhead{K21-2} & \colhead{N24} & \colhead{E26-1}}
\startdata
Fixed $\alpha$ & 2.000 & 2.500 & 1.300 & 1.000 \\
Fixed $D_{\rm exp}$ (km) & 8.23e+01 & 1.30e+02 & 2.58e+01 & 6.29e+01 \\
\hline
Observed $\beta$ & $1.260^{+0.360}_{-0.480}$ & $1.770^{+0.390}_{-0.810}$ & $0.950^{+0.400}_{-0.300}$ & $0.900^{+0.500}_{-0.500}$ \\
\Fiducial & $\bmath{0.826^{+0.040}_{-0.039}}$ \scalebox{0.8}{(0.062)} & $0.431^{+0.100}_{-0.092}$ \scalebox{0.8}{(0.073)} & $\bmath{0.657^{+0.011}_{-0.011}}$ \scalebox{0.8}{(0.042)} & $\bmath{1.142^{+0.019}_{-0.020}}$ \scalebox{0.8}{(0.013)} \\
\LowZ & $\bmath{1.048^{+0.091}_{-0.097}}$ \scalebox{0.8}{(0.042)} & $\bmath{1.263^{+0.147}_{-0.169}}$ \scalebox{0.8}{(0.043)} & $\bmath{0.745^{+0.043}_{-0.045}}$ \scalebox{0.8}{(0.045)} & $\bmath{1.048^{+0.062}_{-0.064}}$ \scalebox{0.8}{(0.047)} \\
\LowG & $0.731^{+0.039}_{-0.041}$ \scalebox{0.8}{(0.107)} & $0.288^{+0.095}_{-0.088}$ \scalebox{0.8}{(0.108)} & $0.613^{+0.012}_{-0.012}$ \scalebox{0.8}{(0.087)} & $\bmath{1.029^{+0.019}_{-0.020}}$ \scalebox{0.8}{(0.066)} \\
\HighP & $-0.579^{+0.014}_{-0.014}$ \scalebox{0.8}{(0.068)} & $-0.680^{+0.007}_{-0.007}$ \scalebox{0.8}{(0.050)} & $0.218^{+0.017}_{-0.018}$ \scalebox{0.8}{(0.127)} & $\bmath{0.626^{+0.020}_{-0.020}}$ \scalebox{0.8}{(0.121)} \\
\LowT & $0.647^{+0.108}_{-0.122}$ \scalebox{0.8}{(0.195)} & $0.522^{+0.220}_{-0.266}$ \scalebox{0.8}{(0.202)} & $0.560^{+0.044}_{-0.049}$ \scalebox{0.8}{(0.183)} & $\bmath{0.820^{+0.058}_{-0.066}}$ \scalebox{0.8}{(0.175)} \\
\Gole & $\bmath{1.055^{+0.050}_{-0.049}}$ \scalebox{0.8}{(0.019)} & $\bmath{1.239^{+0.108}_{-0.104}}$ \scalebox{0.8}{(0.020)} & $\bmath{0.739^{+0.017}_{-0.019}}$ \scalebox{0.8}{(0.017)} & $\bmath{1.141^{+0.028}_{-0.030}}$ \scalebox{0.8}{(0.040)} \\
\enddata
    \tablecomments{ The VTPL parameters from each observational study, and the fitted $\beta$ parameter to each predicted size distribution using $\alpha$ and \D\ constrained to their observationally inferred values, with the 0.84 and 0.16 quantiles as upper and lower bounds.  The \D\ from observation are calculated from the given $H_B$ using the assumptions from Section~\ref{sec:obs_unc} and Section~\ref{sec:disk_model}. The power law exponents ($\alpha$, $\beta$) from the observational data are multiplied by a factor of 5 to account for the change from $H$ to size space as explained in Section~\ref{sec:vtpl}.  For each table entry, those in which the upper and lower bounds on $\beta$ overlap with the quoted range of $\beta$ values from observation are in bold, and a one-sample K-S test statistic $D_{\rm KS}$ between the PSD and fitted VTPL is reported in parentheses to quantify the distributional discrepancy.  When $\alpha$ is lower, there are more instances of agreement between $\beta$ values from observations and those from the PSDs.
    }
    \label{tab:constrained_Hill_fits}
\end{deluxetable*}

\begin{deluxetable*}{cccc}[htb!]
\tablecaption{VTPL fits to $\beta$ alone (continued)}
\tablehead{ & \colhead{E26-2} & \colhead{B26-1} & \colhead{B26-2}}
\startdata
Fixed $\alpha$ & 1.050 & 0.625 & 1.465 \\
Fixed $D_{\rm exp}$ (km) & 4.37e+01 & 1.40e+01 & 6.07e+01 \\
\hline
Observed $\beta$ & $0.850^{+0.500}_{-0.500}$ & $0.745^{+0.305}_{-0.245}$ & $1.205^{+0.305}_{-0.245}$ \\
\Fiducial & $\bmath{0.899^{+0.013}_{-0.014}}$ \scalebox{0.8}{(0.025)} & $\bmath{0.645^{+0.006}_{-0.006}}$ \scalebox{0.8}{(0.027)} & $0.933^{+0.022}_{-0.021}$ \scalebox{0.8}{(0.038)} \\
\LowZ & $\bmath{0.901^{+0.052}_{-0.053}}$ \scalebox{0.8}{(0.044)} & $\bmath{0.676^{+0.030}_{-0.032}}$ \scalebox{0.8}{(0.045)} & $\bmath{0.983^{+0.064}_{-0.070}}$ \scalebox{0.8}{(0.043)} \\
\LowG & $\bmath{0.824^{+0.015}_{-0.014}}$ \scalebox{0.8}{(0.075)} & $\bmath{0.609^{+0.007}_{-0.007}}$ \scalebox{0.8}{(0.072)} & $0.845^{+0.022}_{-0.022}$ \scalebox{0.8}{(0.088)} \\
\HighP & $\bmath{0.469^{+0.015}_{-0.015}}$ \scalebox{0.8}{(0.124)} & $0.454^{+0.006}_{-0.006}$ \scalebox{0.8}{(0.098)} & $0.026^{+0.031}_{-0.030}$ \scalebox{0.8}{(0.115)} \\
\LowT & $\bmath{0.701^{+0.049}_{-0.053}}$ \scalebox{0.8}{(0.177)} & $\bmath{0.553^{+0.027}_{-0.028}}$ \scalebox{0.8}{(0.173)} & $0.713^{+0.068}_{-0.075}$ \scalebox{0.8}{(0.185)} \\
\Gole & $\bmath{0.941^{+0.021}_{-0.022}}$ \scalebox{0.8}{(0.027)} & $\bmath{0.680^{+0.011}_{-0.011}}$ \scalebox{0.8}{(0.018)} & $\bmath{1.018^{+0.030}_{-0.032}}$ \scalebox{0.8}{(0.024)} \\
\enddata
\tablecomments{Continued table entries for Table~\ref{tab:constrained_Hill_fits}}
\label{tab:constrained_Hill_fits_cont}
\end{deluxetable*}

\begin{figure*}[htb!]
    \centering
    \includegraphics[width=\linewidth]{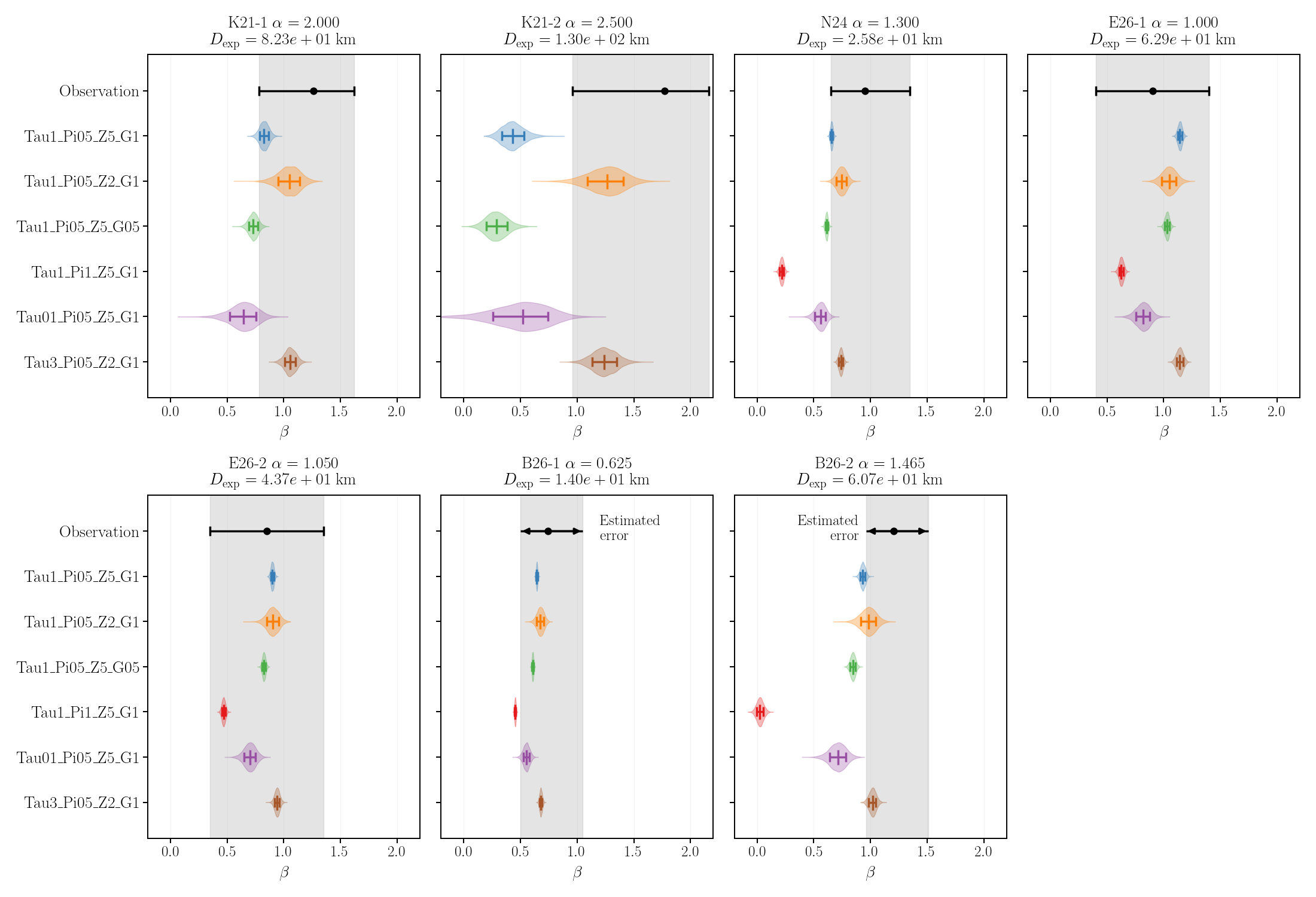}
    \caption{ Violin plots of the posterior distribution of the $\beta$ parameter (x-axis) fitted to each PSD using the observational constraints given by the values in each panel's title. For the first two panels, the \HighP\ simulation's posterior distribution contains all negative values, and is therefore outside the viewing window. The tapered power law parameters presented in \citet{Bernstein2026} are not listed with uncertainty values. However, they present a scatterplot of samples for fitted $\beta_{H_r}$ vs. $\alpha_{H_r}$ (for their generalized $\Gamma$ distribution of the \textit{bright-app} sample, Figure 5 of \citealt{Bernstein2026}), which we use to visually estimate the range on $\beta$ for the B26 panels. } 
    \label{fig:constrained_Hill_fits}
\end{figure*}

We see that using the $\alpha$ and \D\ from observations is not always favored by the simulations. Specifically, the \LowG\ and \HighP\ simulations never have a low distributional discrepancy, indicating they are not able to be described well by the observational values of $\alpha$ and \D, regardless of $\beta$ value. For those that have a low distributional discrepancy, we find agreement between the $\beta$ values fit to the PSDs and observational distributions (e.g. \LowZ). 

There is an apparent trend where PSD $\beta$ values agree with observations more often when the observationally constrained $\alpha$ values are lower. For instance, the largest $\alpha$ (K21-2) has the smallest number of PSDs that show reasonable agreement with observations, while the significantly lower observationally-fit $\alpha$ values (E26-1, E26-2, B26-1) have a broader agreement. Note that no parameter ranges for the VTPL are quoted in \citet{Bernstein2026}. However, the $\beta$ values fit to the PSDs are much closer to the value in B26-1 (lower $\alpha$) than B26-2 (higher $\alpha$). In addition, the range for $\beta$ from \citet{Bernstein2026} can be estimated visually from their scatter plot for fitted $\beta$ vs $\alpha$, this shows that the $\beta$ values fit to the PSDs do generally fall within the range from B26-1 (see Figure~\ref{fig:constrained_Hill_fits}). Altogether, this is suggesting when the observationally fit $\alpha$ values are lower, there is stronger agreement between the observed and predicted size distributions.

In summary, it is clear that the PSDs for a number of runs are demonstrably similar to observation, especially when fixing $\alpha$ and \D\ to the observationally constrained values. This agreement is supported by the K-S tests (Section~\ref{sec:first_comp}) and the consistent agreement between the $\beta$ values fit to observations and the PSDs. The SI is capable of producing the correct exponential taper in the CCKB size distribution.


\section{Discussion} \label{sec:uncertainties}

\subsection{Implications for the Formation History of CCKBOs}

If the SI is the correct model for CCKBO formation, the parameter study presented here can be used to place tentative constraints on the initial conditions of the CCKB by identifying which simulations best reproduce the observed CCKBO size distribution. In the sampled K-S analysis (Section~\ref{sec:first_comp}), \LowT\ shows the strongest agreement with the observations, whereas the parametrized analysis (Section~\ref{sec:vtpl}) most strongly favors \LowZ\ and \Gole. The \Gole\ simulation is favored in more of the parametrized comparisons than \LowT\ --- although both are less favored than \LowZ\ --- whereas \Gole\ is not favored by the sampled analysis. This difference suggests that the parametrized analysis may be more sensitive to changes in $Z$, while the sampled analysis may be more sensitive to changes in $\tau$. Taken together, these results suggest that the CCKB may be most consistent with formation conditions characterized by $\tau \lesssim 0.1$, and more strongly, $Z<0.05$. The constraint on $\tau$ is less clear, however, because the $\tau=0.3$ performs well in the parametrized analysis, but is unremarkable in the sampled comparison. Finally, the results favor $Z=0.02$ over the fiducial $Z=0.05$, although a solar value of $Z=0.01$ was not examined. 

Since the \LowG\ simulation exhibits a minor favorability in the sampled comparison, but a poor favorability in the parametrized comparison, it is unclear if PSDs suggest a primordial CCKB with $\tilde{G}>0.1$ or $\tilde{G} < 0.1$. Finally, \HighP\ does not show strong agreement with the CCKB in either analysis, suggesting that $\Pi<0.1$ is preferred.

These inferred constraints should be regarded as tentative. Thus far, each simulation varies only a single parameter from the fiducial parameter combination. Future work should therefore test a broader region of parameter space to determine, for example, whether simulations with $\tau<0.1$ consistently produce stronger agreement between the simulated PSDs and the observed CCKB size distribution, or whether this agreement depends on the values of the other dimensionless parameters. We also emphasize that interpreting these trends as constraints on the actual conditions of the primordial CCKB requires the assumption that the SI is the mechanism responsible for CCKBO formation.

\subsection{Comparison to Previous IMF Studies}

An additional result from our work is that the IMF from SI simulations is not universal; we find that any physical parameter change (within the bounds of that which we probed) has noticeable effects on the resulting IMF/PSD, so much so that statistical comparisons with the CCKB size distribution are affected. These results reinforce the interpretation that the planetesimal IMF is set by the local disk environment—rather than representing a universal outcome of the instability.

This finding agrees with \cite{Li_2019} but in stands in contrast to earlier SI work focused on the IMF \citep{Simon_2016,Simon_2017}. However, these earlier studies used a different clump finding algorithm, which tended to limit the mass range of detected clumps. They also used individual snapshots to characterize the IMF instead of a clump-tracking algorithm as we have done here. We suspect that the main reason for the different conclusions results from the different clump finding algorithm since \cite{Li_2019} also characterized the IMFs from individual snapshots.

\subsection{Simulation Uncertainties}


\subsubsection{Stochasticity} \label{sec:unc_sto}
A source of uncertainty in our results is the stochasticity associated with a given simulation and the resulting PSD.  \cite{2021MNRAS.500..520R} ran simulations with different seeding of the random perturbations to the system while fixing the parameter values; they demonstrated a degree of stochasticity in calculated planetesimal IMFs (though these were not using a clump tracking algorithm). These simulations were performed at a resolution of 600 grid zones per $H$ for varying grid domain sizes.  Here, we carry out a test of the stochasticity of our PSDs by repeating a simulation with three different random seeds for particle location and velocity fluctuations (as described in Section~\ref{sec:methods}), and comparing the resulting PSDs. 

\begin{figure*}[htb!]
    \centering
    \includegraphics[width=0.9\textwidth]{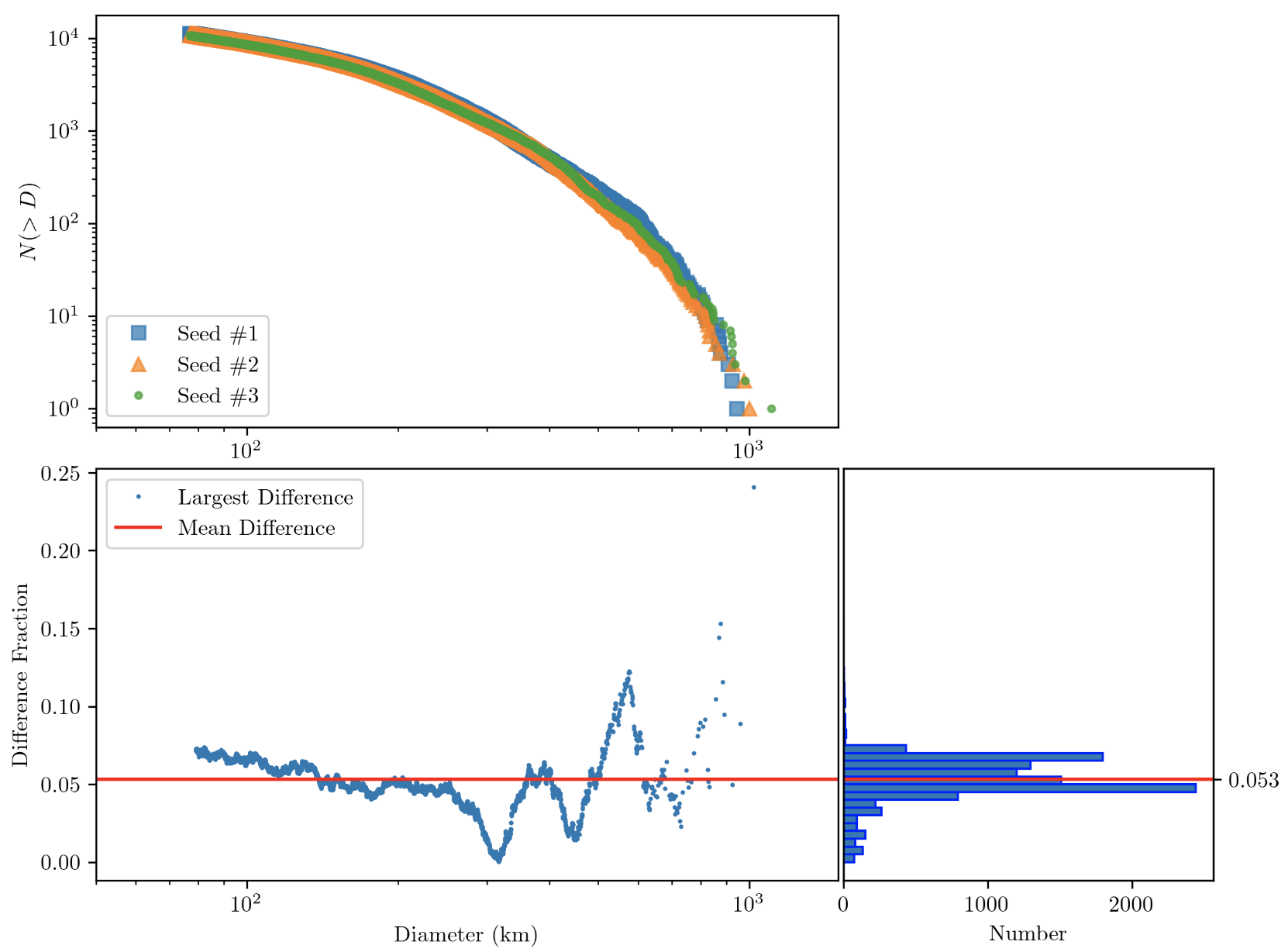}
    \caption{Top: The size distributions of the PSDs for the three randomly seeded runs. The total planetesimal number has not been scaled at all, and is thus different for each simulation. Bottom left: the largest difference between the three PSDs presented above as a function of diameter, expressed as a fraction of diameter. The average of these differences is the horizontal red line. Bottom right: a histogram of the fractional differences in size for the number of objects in the PSDs; most differences are at about 5\%. The differences are smaller at smaller diameters.}
    \label{fig:stochastic_effects}
\end{figure*}

The parameters chosen were those of the \Fiducial\ runs. Figure~\ref{fig:stochastic_effects} shows the PSDs corresponding to each run. The final diameters contained in the PSDs for each seed differed by approximately 5\%  except in the case of the largest diameter (which differed by $\sim25\%$). This uncertainty is smaller than the uncertainty on disk models, albedos, and observational debiasing which will be explored more in Sections~\ref{sec:unc_disk} and Section~\ref{sec:obs_unc}. Additionally, a two-sample K-S test between each of the three random seeds yield $p$-values greater than 0.6 for all three combinations. Based on this experiment, we believe that the PSD shapes is robust to small changes in the initial conditions, and it is appropriate to take as representative one simulation from each set of parameters.

One way to reconcile this result with the results of \cite{2021MNRAS.500..520R} is by considering the differences in the numerical resolution. Our fiducial resolution for these simulations is 2560 $H^{-1}$, whereas \cite{2021MNRAS.500..520R} employed a resolution of 600 $H^{-1}$. \cite{2021MNRAS.500..520R} reports a spread in planetesimal numbers from 14 to 22 in their smallest domain. In order to estimate what the IMFs in the simulations in this study would look like had they been performed at 600 $H^{-1}$, we can look at the number that had diameters larger than $2560/600 \times D_{\rm min}$ for the runs. This corresponds to a spread of 935 to 969 objects. We believe our numbers were much greater for three reasons. First, the physical parameters of the \Fiducial\ run are more conducive to planetesimal formation via the SI than those of \citet{2021MNRAS.500..520R} (namely $Z=0.05$ vs $Z=0.02$). Second, the smallest domain size of \cite{2021MNRAS.500..520R} is half the volume of the \Fiducial\ run, which could easily result in fewer planetesimals. Third, \citet{2021MNRAS.500..520R} performs mass distribution analysis at a single time snapshot in the simulation, potentially missing objects that formed and subsequently were destroyed/merged at earlier times; our clump-tracking algorithm accounts for this, as outlined in Section~\ref{sec:plan_ct}, potentially leading to a larger number of total planetesimals formed.

Thus, while it is still possible that stochasticity ultimately can affect the IMF, the increased planetesimal numbers and advanced clump tracking in our simulations compared with \citet{2021MNRAS.500..520R} suggest that the impact of this effect is negligible. That said, it should be kept in mind that we only tested the effect of stochasticity for one combination of physical parameters, and extending this testing to other parameter combinations is left to future work.

\subsubsection{Simulation Integration Time} \label{sec:unc_len}

The length of time over which the simulation is run may also have an impact on the PSDs. While {\sc ClumpTracker} quantifies the masses of planetesimals as soon as they form, as described above, these planetesimals are not being removed from the simulation. As a result, there is no way to completely rule out the effects of already-formed planetesimals on the later stage evolution of the system. For example, these planetesimals, which are more like extended pebble clouds (as discussed above), might accrete solids from the surrounding environment at a higher rate than if they were fully collapsed objects. This increased accretion of pebbles will not be combined into the planetesimal mass in the IMF, but may rob the local environment of pebbles that would otherwise form additional planetesimals. In short, our PSDs may be not completely reflect the true PSD produced by the SI. Given this limitation, a shorter simulation runtime should be prioritized. 

Conversely, not all shearing box simulations produce planetesimals at the same rate, even after the activation of self-gravity. In order to better enable a robust comparison between observations and our predictions, our simulations must generate a sufficiently large population of planetesimals (see Section~\ref{sec:vtpl}). Therefore, we must consider both effects: prolonged evolution that may distort the true IMF and run times too short to enable a good comparison with observations. 

Thus, we must quantify the effect of the planetesimal formation time on the IMF/PSD.
In order to do this, we run the {\sc ClumpTracker} algorithm from the start of planetesimal formation to some end time; we then increase that end time and determine whether the PSD changes in a significant way. In Figure~\ref{fig:multiclumpt}, we chart the inter-quartile range (IQR) of the \Fiducial\ PSD over different {\sc ClumpTracker} end times, with the actual PSD distribution shown adjacent. If we were to perform a random selection of an unknown underlying distribution over time,  the resulting sample should converge in appearance, represented by quartiles and a median approaching a constant value as enough objects are sampled. 

The onset of convergence is evaluated by measuring how rapidly the PSD changes over increasing {\sc ClumpTracker} end times. Rather than using a traditional time derivative, we assess the fractional change in the IQR diameter values between successive {\sc ClumpTracker} end times.  Thus, Figure~\ref{fig:multiclumpt} also shows the IQR and the fractional change over time in the IQR for the \Fiducial\ run. 

This simulation continued for only 10$\Omega^{-1}$ after the onset of self-gravity, yet the IQR converged in value quickly: within 5$\Omega^{-1}$, the fractional changes dropped below 5\%, and by 7$\Omega^{-1}$, they fell below 3\%. To ensure comparable conditions across all runs, each simulation was continued until the fractional change in size remained below 3\% for at least 20\% of the time elapsed since the activation of self-gravity. 
For instance, in the \Fiducial\ run, the  threshold was reached at approximately 6.5$\Omega^{-1}$ after the activation of self-gravity. Since the simulation concluded at 10$\Omega^{-1}$, the threshold was maintained for about 35\% of the post-activation runtime. The \LowT\ run reached the threshold at roughly 75$\Omega^{-1}$ and completed at 96$\Omega^{-1}$, meaning the threshold was satisfied for approximately 22\% of the simulation duration following self-gravity activation. We verified all simulations met this criteria in the same way.

It is essential to evaluate the threshold before removing the low-resolution edge of the IMF/PSD (Appendix~\ref{sec:resolution_effects}). Truncating the PSD before analysis reduces the total number of objects, possibly to the point where minor fluctuations in the distribution can cause brief deviations beyond the 3\% fractional change threshold at specific {\sc ClumpTracker} end times. These deviations are short-lived, resolving within 0.5$\Omega^{-1}$. In simulations such as \LowZ\ and \LowT, enforcing the 3\% threshold based on the PSD with the low-resolution edge removed could result in unnecessarily extended run times, even though the full, non-truncated PSD has already converged for a significant duration.

Encouragingly, in all simulations, once the 3\% fractional change threshold was reached simultaneously for both the IQR and the median, it was never exceeded again. While we cannot definitively rule out the possibility that had we run the simulations further, planetesimals formed at later times would influence the PSD, our results strongly suggest that the late formation of planetesimals does not appreciably change the PSDs used in this study.

\begin{figure*}
    \centering
    \includegraphics[width = 0.9\textwidth]{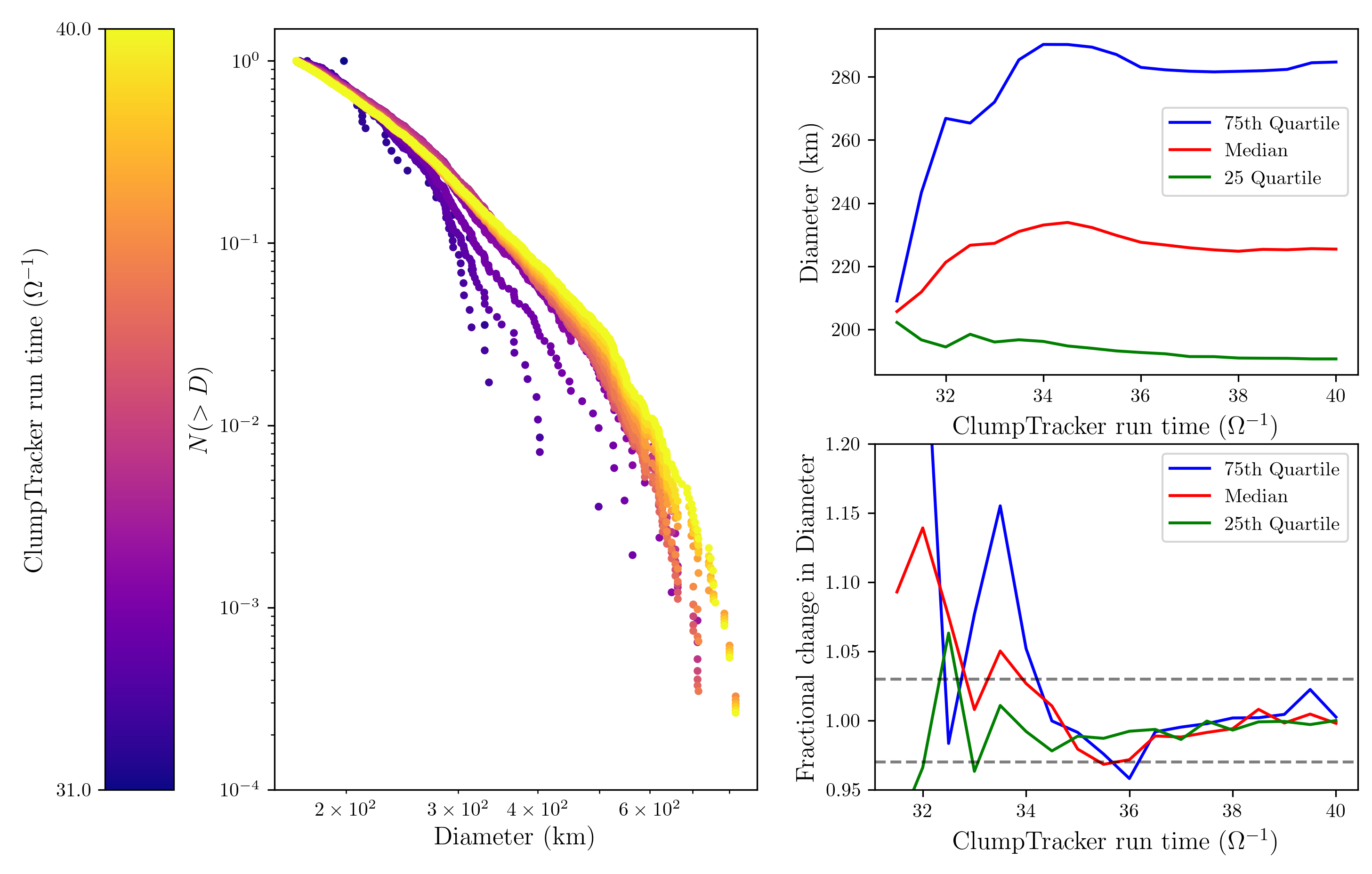}
    \caption{Left: the PSD for the \Fiducial\ run, taking {\sc ClumpTracker} to ever increasing lengths of time, at 0.5$\Omega^{-1}$ time increments from 31 to 40 $\Omega^{-1}$, with the number of planetesimals normalized by the total number over the length of time analyzed. Since the total number of objects in the PSD increases the longer the simulation runs, this allows us to compare the shapes of the PSDs without needing to account for this increase. Upper right: The inter-quartile range (IQR) of the PSD diameters vs time. Lower right: the fractional change of the IQR as a function of {\sc ClumpTracker} end time, with a 3\% range shown by the dashed lines. For sufficiently long run times, the IQR no longer exceeds the 3\% threshold, demonstrating the relatively fast convergence of the shape of the PSD over time.} 
    \label{fig:multiclumpt}
\end{figure*}

\subsubsection{Disk Model} \label{sec:unc_disk}
As mentioned, the value of \D\ for a given PSD depends on the disk model used to convert code to physical units. There are a number of free parameters in our disk model that are not strongly constrained by observations, namely the values of $M_d$, $R_c$, and $q$. For example, by allowing $M_d$ to be as low as $0.01 M_\odot$ or as large as $0.2 M_\odot$, $R_c$ between 30 {\rm AU} and 100 {\rm AU}, and $0.4<q<0.6$, $M_H$ can be as low as $9.57\times 10^{26}$ g or as high as $2.42 \times 10^{28}$ g. That means that the largest clump formed in our fiducial simulation, at $7.73\times 10^{-6}$ code units, could range in diameter from 383 km to 1239 km. For reference, the brightest candidate in recent Cold Classical belt studies is $H_r \approx 3.6$ \citep{largestCCB}; at an albedo of 0.15, this corresponds to a diameter of 653 km.

The unknown disk model parameters is the largest source of uncertainty on the theory side of this work. However, this uncertainty is not expected to affect the PSD shape since it leads to only an uncertainty in the mass scaling used to convert code units to physical units. The disk model choices for the parametrized analysis (Section~\ref{sec:vtpl}, where the analysis depends on a disk model) are presented in Appendix~\ref{app:dm}.


There is a wide range of both disk masses ($M_d$) and characteristic radii ($R_c$) that equate to an $M_H$ that aligns the PSDs with the MPC (see Section~\ref{sec:vtpl}). To assess whether these required $M_H$ values can be produced by plausible Solar System-like disk models, we evaluate two representative parameter combinations for each PSD: a relatively massive disk with $M_d = 0.1\ M_\odot$, from which we solve for $R_c$, and a Solar System-like characteristic radius of $R_c = 45\ {\rm AU}$, from which we solve for $M_d$. We consider a PSD consistent with Solar System-like disk parameters if the $R_c = 45 \ {\rm AU}$ model requires a disk mass above that of the MMSN \citep[$\gtrsim 0.013\ M_\odot$,][]{mmsn1, mmsn2}. Three of six PSDs meet this criterion and can therefore be reproduced with Solar System-like disk parameters. Encouragingly, these are \LowT, which showed strong similarity to the CCKB distribution in Section~\ref{sec:first_comp}, and \LowG\ and \Gole, which showed strong similarity to all observational fits in Section~\ref{sec:vtpl}.



\subsection{Observational Uncertainties of CCKBOs} \label{sec:obs_unc}

The CCKB has been observed to exhibit geometric albedos spanning approximately 0.1–0.2, with reported mean values near $\sim0.16$ based on available samples \citep{2014ApJ...782..100F,2014ApJ...793L...2L}. However, these estimates are derived from limited and potentially selection-biased samples, and the intrinsic albedo distribution remains uncertain. For this reason, it is standard practice to adopt a fiducial albedo of 0.15 when converting magnitudes to sizes. We follow this convention throughout this work, noting that variations across the observed albedo range correspond to an uncertainty of order $\sim 20\%$ in inferred diameters, while the difference between assuming 0.15 and 0.16 is negligible compared to this intrinsic scatter. 

An albedo that varies with different CCKBOs would present a complication that would be challenging to include in this study. Fortunately, the relationship between albedo and size has been studied for different TNO object populations \citep{2012A&A...541A..94V,2018A&A...618A.136V}. Since CCKBOs have very minimal collisional evolution, and have similar surface characteristics as suggested by their spectra \citep{2021PSJ.....2...57S}, it is not expected that they possess a significant albedo distribution beyond a uniform value, so the use of constant albedo is acceptable.

Besides albedo, the bulk material density of CCKBOs could range from 0.30 to 3.74 ${\rm g}/{\rm cm}^3$, corresponding to a diameter difference of $\sim130\%$ (this density consideration does not account for porosity or planetesimal shape). Should density depend on planetesimal mass, or have some other complicated relationship with size, this would also distort the PSD shape. The standard assumption is uniform density of CCKBOs at $\rho_{\rm mat} = 1.0 \ {\rm g}/{\rm cm}^3$. This choice is supported by the observed density of similar size objects \citep{2024ApJ...974....9T}, and while theory shows a spread in material densities \citep[$\sim 0.5 - 1.5 {\rm g}/{\rm cm}^3$ for similar mass objects $\sim 10^{-3}-10^{-4}\ M_{\rm Pluto}$;][]{2024PSJ.....5...55C}, the spread is small. Thus, for simplicity we assume constant density; this should be kept in mind when interpreting results. 


\subsection{Binary Formation Effects}
\label{sec:binary}

The resolution of our simulations prevents us from following the collapse of an SI-induced pebble cloud to actual planetesimal scales \citep{Simon_2016}. Sub-grid collapse and fragmentation into planetesimals requires techniques that are outside the scope of this study \citep{2017ApJ...835..109W, 2020ApJ...901...54K}. Therefore, we assume a monolithic spherical collapse for simplicity. This monolithic collapse assumption is commonly made in the SI literature, but is indeed an additional source of uncertainty in our results. Observations of wide binaries, along with simulations that follow the collapse of pebble clouds, demonstrate that binary formation and even higher-order multiplicity formation are likely \citep{Nesvorny_binary, 2021PSJ.....2...27N, 2025ApJ...990..130B}. Additionally, excess angular momentum can lead to non-spherical, flattened planetesimal shapes, further altering the diameter calculations from a pebble cloud mass \citep{2024A&A...683A..38L, 2025ApJ...990..130B}.

Carrying out similar types of calculations as \cite{2021PSJ.....2...27N}, where individual clumps are extracted from SI simulations is beyond the scope of this work. In lieu of this, we use a simple toy model of collapse to mimic the effect of further collapse on the resulting IMF and PSD. This model, which is based on the results of \cite{2021PSJ.....2...27N}, will use an individual pebble cloud's scaled angular momentum $l$ at the time of formation as: $l = \frac{L}{L_H}$, where $L_H = M R_H^2 \Omega$; M is the mass of a planetesimal, and  $R_H$ is the Hill radius:
\begin{equation}
\label{eqn:hill}
    R_H^3 = \frac{\tilde{G}}{4\pi} \frac{M}{3\Omega},
\end{equation}

For each pebble cloud in the IMF, the following model is applied:
\begin{itemize}
    \item $l < 0.07$: results in a single object, the original mass stays in the IMF as is --- this happens for $\sim 40\%$ of objects. 
    \item $l > 0.07$ and $R < 0.5R_H$: results in a (approximately) equal mass binary. The original mass is removed from the IMF, and two masses with 50\% of the original are added back in to the distribution.
    \item $l > 0.07$ and $R> 0.5R_H$: results in a hierarchal system with some number of satellites. The original mass is removed, and replaced with a single object containing 50\% of the mass of the original. The corresponding satellites are smaller than the resolution scale, so we do not include them in the modified IMF.
\end{itemize}

Using this toy model, we plot  the PSDs for two of our simulations in Figure~\ref{fig:binary}. The final result is a slightly ``steeper'' PSD, as larger masses are converted into multiple smaller bodies, depopulating the high-mass end and repopulating the low mass end. We fit new $\beta$ values to each PSD after the toy collapse model was applied, again using the observationally constrained $\alpha$ and \D. All values of $\beta$ increase, as expected for steeper distributions. However, all $\beta$ values that agreed with observations (to within the observational uncertainties) before the toy model was applied, continued to agree with observations (again within the uncertainties) after application of the model. These considerations suggest that the monolithic collapse assumption has negligible effect on the comparisons with observation. However, a more realistic collapse method could potentially lead to different results. Employing methods to quantify this effect in more detail, such as mesh refinement or zoom-in calculations following \cite{2021PSJ.....2...27N}, are left for future studies.

\begin{figure}[htb!]
    \centering
    \includegraphics[width=1\linewidth]{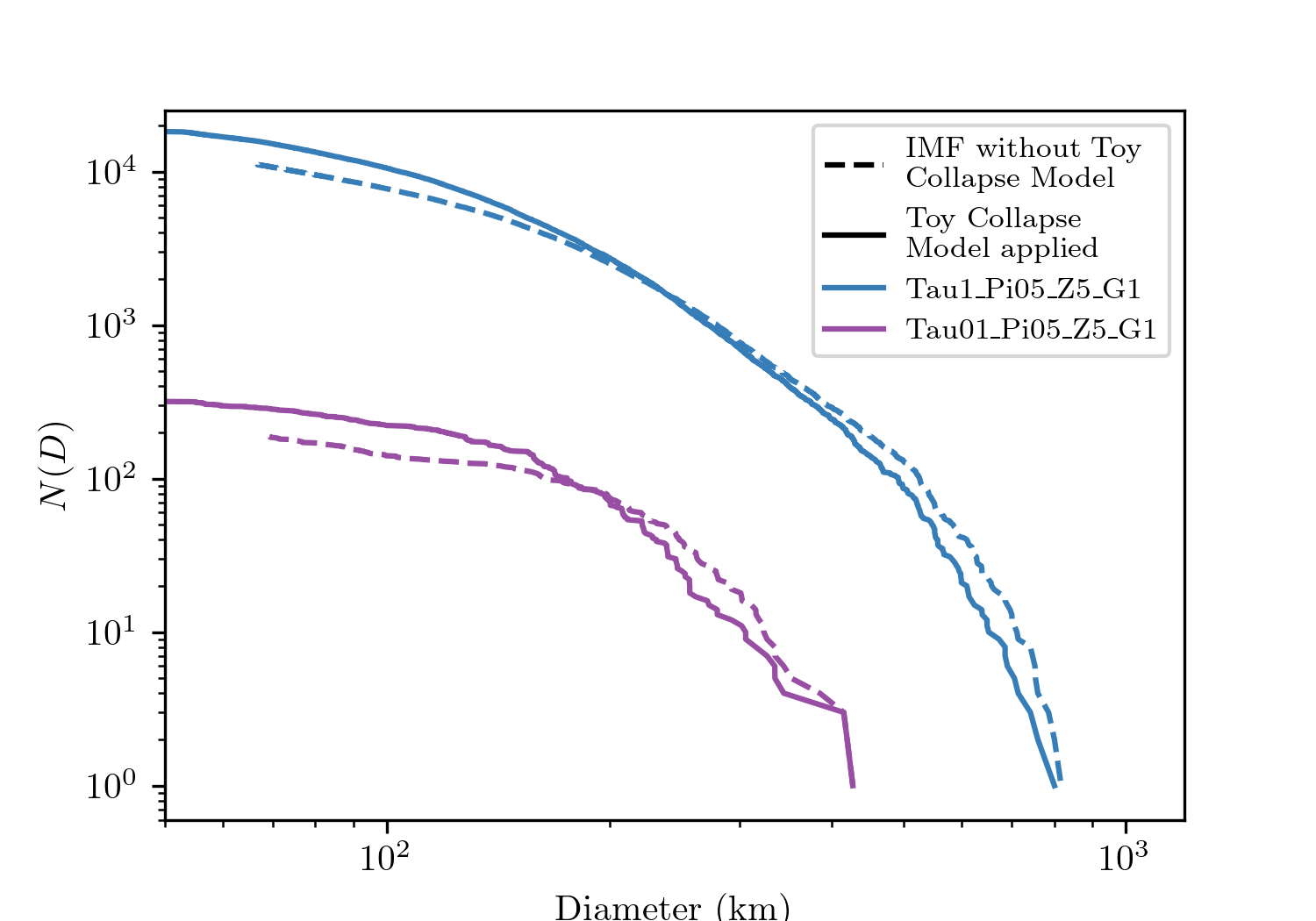}
    \caption{The PSDs from \Fiducial\ and \LowT\ in original form (dashed lines) and after adding our toy binary collapse model (see main text; solid lines).  After applying the model, larger objects are transformed into smaller objects, generally changing the slope of the distribution to be steeper. The differences are not large enough to conclude that the monolithically collapsed size distributions are a poor representation, and these differences are significantly smaller than observational uncertainty.}
    \label{fig:binary}
\end{figure}

Notably, \cite{2025ApJ...990..130B} found that a rotating spherical cloud often collapses into higher multiplicity of objects (dozens to hundreds of planetesimals), depending on the rotation speed of the cloud. Although the clouds in said study were not formed via the SI, we can perform a first-order comparison with our results by making assumptions about the size, shape, and rotations of our clouds. This should determine if the clouds in our study are best represented by slowly-rotating, likely binary collapse (c.f. \citealt{Nesvorny_binary}), or rapidly-rotating, likely fractured higher multiplicity collapse. \cite{2025ApJ...990..130B} defines a rotation rate scale factor $f$ as $\Omega_{\rm cloud} =f\Omega_{\rm circ}$, where $\Omega_{\rm circ} = \left(GM_{\rm cloud}/R_{\rm cloud}^3\right)^{1/2}$ is the circular orbital angular frequency at the edge of their initialized spherical cloud, $R_{\rm cloud} = 0.6R_H$. 

For our simulations, the cloud angular frequency can be approximated from the cloud's $l$, $L_H$, and moment of inertia $I_{\rm cloud}$ as $\Omega_{\rm cloud} = lL_H/I_{\rm cloud}$. For simplicity, $I_{\rm cloud}\approx k_g MR_H^2$, where $k_g$ is a fraction representing the geometry of the cloud, e.g. $k_g=\frac{2}{5}$ for a solid uniform sphere. This simplifies to $\Omega_{\rm cloud}=l\Omega/k_g$. Therefore, had our pebble clouds been in \citeauthor{2025ApJ...990..130B}'s study, we can calculate a corresponding $f$ from the scaled angular momentum $l$ as:
\begin{equation}
    f = \frac{l}{k_g}\left(\frac{R_{\rm cloud}^3}{GM_{\rm cloud}} \right)^{1/2}\Omega.
\end{equation}
$R_{\rm cloud}$ must be approximated, since none of our clumps are spherical by nature. We can define a dimensionless radius fraction $\xi \equiv R_{\rm cloud}/R_H$, that will allow us to approximate the size of a cloud as a sphere with radius a fraction of its Hill radius. For \cite{2025ApJ...990..130B}, they initialize all of their clouds with a uniform density and $\xi = 0.6$. For our simulations, clouds are not uniformly dense, and a radial density profile extends all the way out to $R_H$. Thus, this is not a perfect characterization of our clouds but provides enough degrees of freedom to examine how our clouds would collapse if evolved such as in \cite{2025ApJ...990..130B}. With the inclusion of  $R_H^3= \frac{GM_{\rm cloud}}{3\Omega^2}$, we can simplify the $f$ relationship to $l$ via the cloud radius fraction $\xi$ and cloud geometry $k_g$:
\begin{equation}
    f = l\ \frac{\xi^{3/2}}{3k_g}.
\end{equation}

Using $\xi=0.6$ and a $k_g = 2/5$ (an upper bound, since the mass distribution will be higher in the center of the cloud), the approximate rotation rate scale factors $f$ for clumps in the \Fiducial\ simulation range from 0.3 prograde to 0.15 retrograde. Examining $\xi=1$ (since the mass extending beyond $0.6R_H$ makes a larger $\xi$ likely more appropriate), the approximate rotation rate range extends to 0.65 prograde and 0.36 retrograde. These are relatively small rotations compared to the maximum 1.2 in \cite{2025ApJ...990..130B}, so it is appropriate to assume that our SI-formed pebble clouds will form lower-multiplicity systems when collapsing, closer to the prevalence of equal size binaries measured in \citet{ 2021PSJ.....2...27N}.

The relationship between complex cloud geometry,  angular momentum, and the relatively simple expression for rotation rate scale factor should only be taken as a first order comparison, since many assumptions about cloud body shape and density distributions were made. The key point is that the SI forms spinning clouds that are on the slow rotation end of the parameter space that determines multiplicity. 





\section{Conclusions} \label{sec:conclusions}

In this work, we simulated the planetesimal formation process via the SI and calculated IMFs and size distributions for different parameters. These predicted size distributions were compared to the observational size distributions of Cold-Classical KBOs. Overall, we found similarities between the simulated and observed size distributions.
 
The main conclusions are as follows:
\begin{itemize}
    \item The SI is able to produce a statistically similar shape to the distribution of the largest objects in the CCKB (those within the exponential taper of the distribution) as evidenced by K-S test results.
    \item The predicted size distributions produced by the SI contain only large planetesimals in the exponentially tapered region of the CCKB distribution. Therefore the small-size end of the distribution (dominated by $\alpha$) and the characteristic tapering size (\D\ or $H_B$) of the CCKB are difficult to probe when fitting. Instead, by holding $\alpha$ and \D\ to observationally fit values and fitting $\beta$ to the PSDs alone, the SI is able to reproduce the $\beta$ value from observations of the CCKB. The agreement in $\beta$ values is best for low $\alpha$ values: $\alpha \lesssim 1.5$, $\alpha_M \lesssim 0.5$, $\alpha_{H_r} \lesssim 0.3$.
    
    \item The predicted size distributions show stronger agreement with observations for $\tau \le 0.1$, $Z<0.05$ (favoring $Z=0.02$ in this study, though solar $Z=0.01$ was not tested),  and $\Pi < 0.1$.
    
    \item The size distributions of planetesimals are robust to small differences in initial conditions with the same physical parameters, when adequate resolution is reached.
    
    \item A toy model including the effect of sub-grid collapse and multiplicity on the size distribution suggests that while the IMF will be steepened by sub-grid collapse, the effect is sufficiently weak such that other uncertainties (e.g., observational) will dominate. 
\end{itemize}

Our results clearly demonstrate that, modulo uncertainties, the size-frequency distributions produced by the SI are in broad agreement with observations of the CCKB for the largest planetesimals. More specifically, for various parameter combinations we explored, we found agreement between our predictions and observations quantified by at least one, and in some cases multiple, statistical tests. While there are cases in which observations and predictions disagree, our results show that the SI can reproduce the observed size distribution of the CCKB within the range of parameters we have explored.  Coupled with previous theoretical and numerical support for the SI from studies of binary collapse \citep{2021PSJ.....2...27N}, our findings support the hypothesis that the SI is the dominant mechanism for forming the CCKB. 
Future work should compare the small planetesimal limits (i.e., $D < 100$ km for the disk model used here) of the predicted and observed CCKB distributions and should compare simulations to other planetesimal populations where possible. However, this work puts the SI on even firmer footing as a key process in outer Solar System planetesimal formation.

%

\begin{acknowledgments}
     We thank J.J. Kavelaars for providing the debiased data used in his 2021 study for direct comparison. WH also acknowledges Tabassum Tanvir and David Rea for their very productive conversations about the role of simulation length on the initial mass function. We thank Hal Levison, Seth Jacobson, and Jackson Barnes for their productive conversations regarding sub-grid collapse of pebble clouds. WH and JBS acknowledge support from the NASA Emerging Worlds program under Grant \# 80NSSC25K7398. J.L. acknowledges support from NASA under the Future Investigators in NASA Earth and Space Science and Technology grant \# 80NSSC22K1322. DC acknowledges support from NASA under Emerging Worlds through grant 80NSSC25K7414 and Exoplanets Research Program through grant 80NSSC24K0959. This work was supported by the NASA High-End Computing (HEC) Program through the NASA Advanced Supercomputing (NAS) Division at Ames Research Center, using the Pleiades supercomputer.
\end{acknowledgments}

\vspace{5mm}


\software{Athena (\url{github.com/PrincetonUniversity/athena}),  
          PLAN (\url{github.com/astroboylrx/PLAN}),
          ClumpTracker (\url{github.com/simon-research-group/clumptracker}),
          emcee (\url{github.com/dfm/emcee}),
          numpy (\url{numpy.org}),
          scipy (\url{scipy.org})
          }


\appendix
\setcounter{table}{0}
\renewcommand{\thetable}{A\arabic{table}}

\setcounter{figure}{0}
\renewcommand{\thefigure}{A\arabic{figure}}

\section{Resolution Effects} \label{sec:resolution_effects}
 
The effect of resolution on IMF has been previously studied \citep{2015SciA....1E0109J,Simon_2016} finding that smaller bodies are formed as the resolution is increased. As is the case with numerical simulations, physical processes near the grid scale are not captured as accurately.

In the CDF form of an IMF/PSD, we can characterize where the distribution sizes become ``poorly-resolved" based on the considerations discussed above by comparing the Hill radii of the smallest planetesimals with the grid cell size. Here, we do this for the \Fiducial, \LowR, and \UltraLowR\ runs. We expect there to be a Hill radius length corresponding to a volume of grid cells in which the size distribution appears to start to become deficient --- representing either a lack of formation or inability to detect objects.

For convenience, we write down the Hill radius again: 
\begin{equation}
    R_H^3 = \frac{\tilde{G}}{4\pi} \frac{M}{3\Omega},
\end{equation}
and the grid cell size $\Delta x$ can be calculated as
\begin{equation}
    \Delta x = \frac{L_x}{N_x} = \frac{L_y}{N_y} = \frac{L_z}{N_z}
\end{equation}
which corresponds to $\Delta x = 0.4 H/1024 = 3.9\times 10^{-4} H$ for the highest resolution fiducial run. The cumulative distributions of the Hill-radii for the three simulations analyzed here are shown in Figure~\ref{fig:hill_radius}.

We aim to determine at what scale resolution effects become dominant. In order to do so, we fit the broken cumulative power law \citep[BCPL;][]{Li_2019}, defined with three parameters, $\alpha_1$, $\alpha_2$, and $R_{\rm br}$ \citep[c.f. $M_{\rm br}$ in][]{Li_2019}:
\begin{equation}
    P(>R) = \begin{cases}
        c_1R^{-\alpha_1} & R \le R_{\rm br} \\
        c_2R^{-\alpha_2} & R > R_{\rm br}
    \end{cases}
\end{equation}
with normalization as
\begin{equation}
    c_1 = \frac{1}{R_{\rm min}}
\end{equation}
\begin{equation}
    c_2 = \frac{1}{R_{\rm min}^{-\alpha_1}R_{\rm br}^{\alpha_1-\alpha_2}}.
\end{equation}

In doing so, we notice a ``plateau'' at small sizes in the distributions. This plateau occurs consistently at $R_H = 2\ \Delta x$, suggesting that it could be the minimum resolution effect. Therefore, we truncate all size distributions to only the objects with $R_H > 2.5 \ \Delta x$.

\begin{figure}[htb!]
    \centering
    \includegraphics[width=\linewidth]{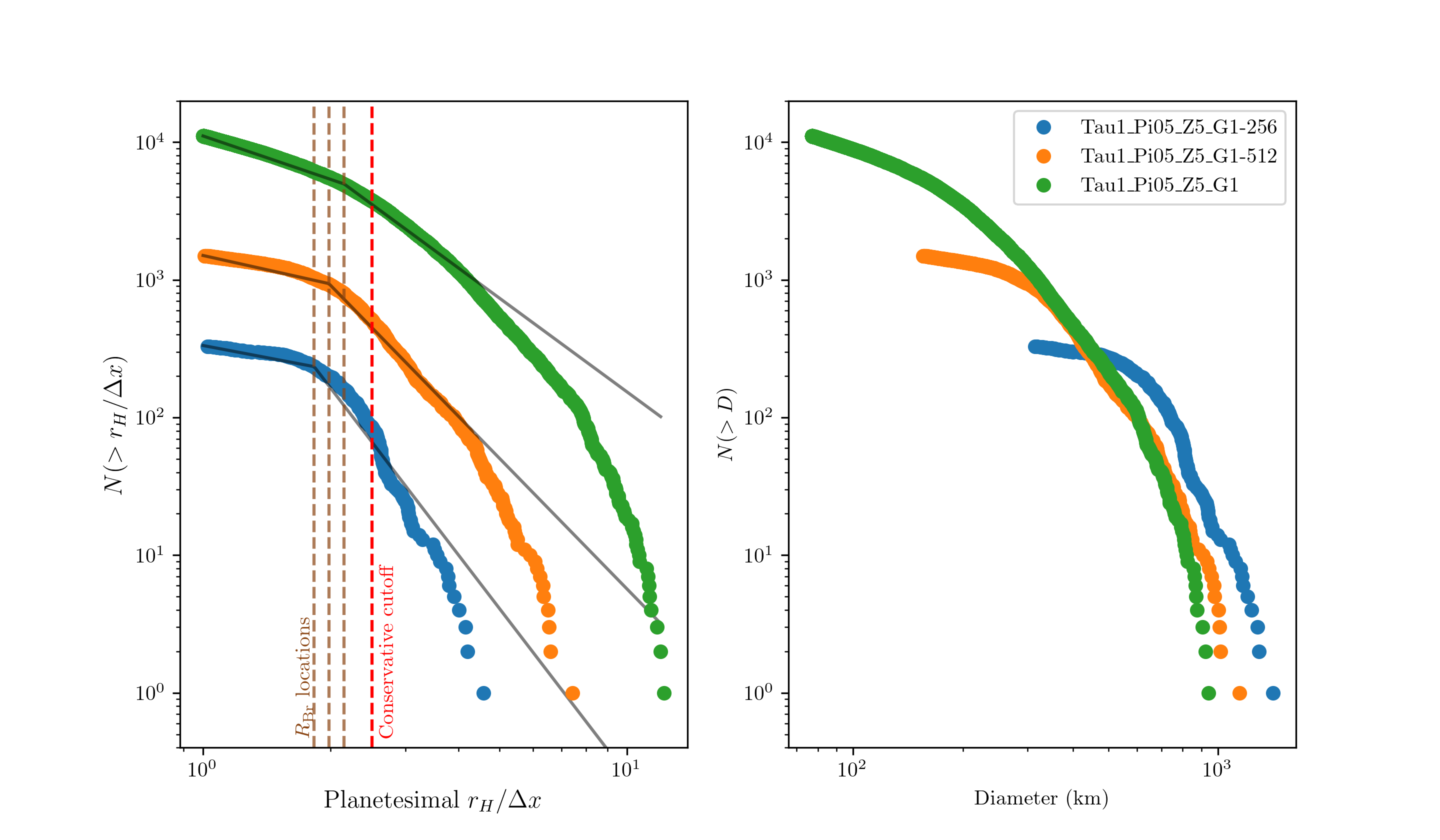}
    \caption{Left: The cumulative size distributions of Hill radii for \Fiducial\ (green), \LowR\ (orange), and \UltraLowR\ (blue), along with BCPL fits to each (black lines), with the corresponding $R_{\rm br}$ locations (orange dashed lines). Right: The PSD of each simulation, where the size-dependent turnovers are visible on top of one another. The red line denotes the cutoff size below which smaller objects are removed as these objects will be less likely to be accurately modelled.}
    \label{fig:hill_radius}
\end{figure}

\section{Disk model values}
\label{app:dm}
In Section~\ref{sec:vtpl}, the masses of the PSD are scaled by aligning the 29th largest body in the PSD with the smallest object of the MPC sample. This scaling assumes an albedo of 0.15 and $\rho_{\rm mat}=1\ {\rm g/cm}^3$.  The appropriate $M_H$ (see Equation~\ref{eq:MH}) to convert from code units to physical units to meet this scaling can then be calculated simply as 
\begin{equation}
    M_H = \frac{\pi \rho_{\rm mat}D_{\rm MPC, 29}^3 /6}{M_{\rm code,29}}, 
\end{equation}
where $D_{\rm MPC, 29}$ is the size of the smallest object in the MPC sample, and $M_{\rm code, 29}$ is the mass of the 29th largest object of the PSD. However, the corresponding $M_d$, $R_c$, and $q$ that equate to this $M_H$ via Equation~\ref{eq:MH} are degenerate; multiple parameter combinations lead to the same value. This is illustrated in Figure~\ref{fig:MH_scaling}, where the PSDs are aligned with observations, and the right panel shows contour lines of constant $M_H$ plotted for varying $M_d$ and $R_c$.

Since $M_H$ depends on $M_d$ and $R_c$, we can explore disk model parameters capable of producing the precise $M_H$ values required to scale each aligned PSD with the MPC sample. For each PSD, we consider two representative parameter combinations. In the first, we fix $R_c = 45 \ {\rm AU}$ and $q=0.5$ and solve for the disk mass $M_d$ required to reproduce the necessary $M_H$.  In the second, we fix $M_d = 0.1\ M_{\odot}$ and $q=0.5$ and solve for the $R_c$ that produces the necessary $M_H$. These parameter combinations are presented in Table~\ref{tab:model_values}.

\begin{deluxetable}{cccc}[htb!]
\tablecaption{$M_H$ model values}
    \tablehead{ Simulation & $M_d$, $R_c$ ($M_\odot$, AU) & $q$ & $M_H$ ($g$)}
\startdata
\multirow{2}{*}{\Fiducial}
    & 0.003, 45  & \multirow{2}{*}{0.5} & \multirow{2}{*}{3.670e+26}\\
    & 0.100, 7.0 &                      &                         \\[2pt]
\multirow{2}{*}{\LowZ}
    & 0.081, 45   & \multirow{2}{*}{0.5} & \multirow{2}{*}{1.030e+28}\\
    & 0.100, 24.9 &                      &                         \\[2pt]
\multirow{2}{*}{\LowG}
    & 0.005, 45  & \multirow{2}{*}{0.5} & \multirow{2}{*}{6.673e+26}\\
    & 0.100, 7.9 &                      &                         \\[2pt]
\multirow{2}{*}{\HighP}
    & 0.001, 45  & \multirow{2}{*}{0.5} & \multirow{2}{*}{8.717e+25}\\
    & 0.100, 5.6 &                      &                         \\[2pt]
\multirow{2}{*}{\LowT}
    & 0.038, 45   & \multirow{2}{*}{0.5} & \multirow{2}{*}{4.839e+27}\\
    & 0.100, 14.5 &                      &                         \\[2pt]
\multirow{2}{*}{\Gole}
    & 0.012, 45  & \multirow{2}{*}{0.5} & \multirow{2}{*}{1.584e+27}\\
    & 0.100, 9.7 &                      &                         \\
\enddata
    \tablecomments{The value of $M_H$ by which to scale each simulation to align with the complete Minor Planet Center sample depends on the adopted disk parameters $M_d$, $R_c$, and $q$. Because $M_d$ and $R_c$ are degenerate in determining $M_H$, we list two representative parameter combinations for each simulation: one with $R_c=45\,\mathrm{AU}$ and one with $M_d=0.1\,M_\odot$. Both combinations yield the same $M_H$. For the PSDs that can be aligned using parameter values consistent with Solar System models \citep{mmsn1, mmsn2, 2012AJ....143...91K, solar_nebula}, we find good agreement with the CCKB size distribution.
    }
    \label{tab:model_values}
\end{deluxetable}

\begin{figure}[htb!]
    \centering
    \includegraphics[width=\linewidth]{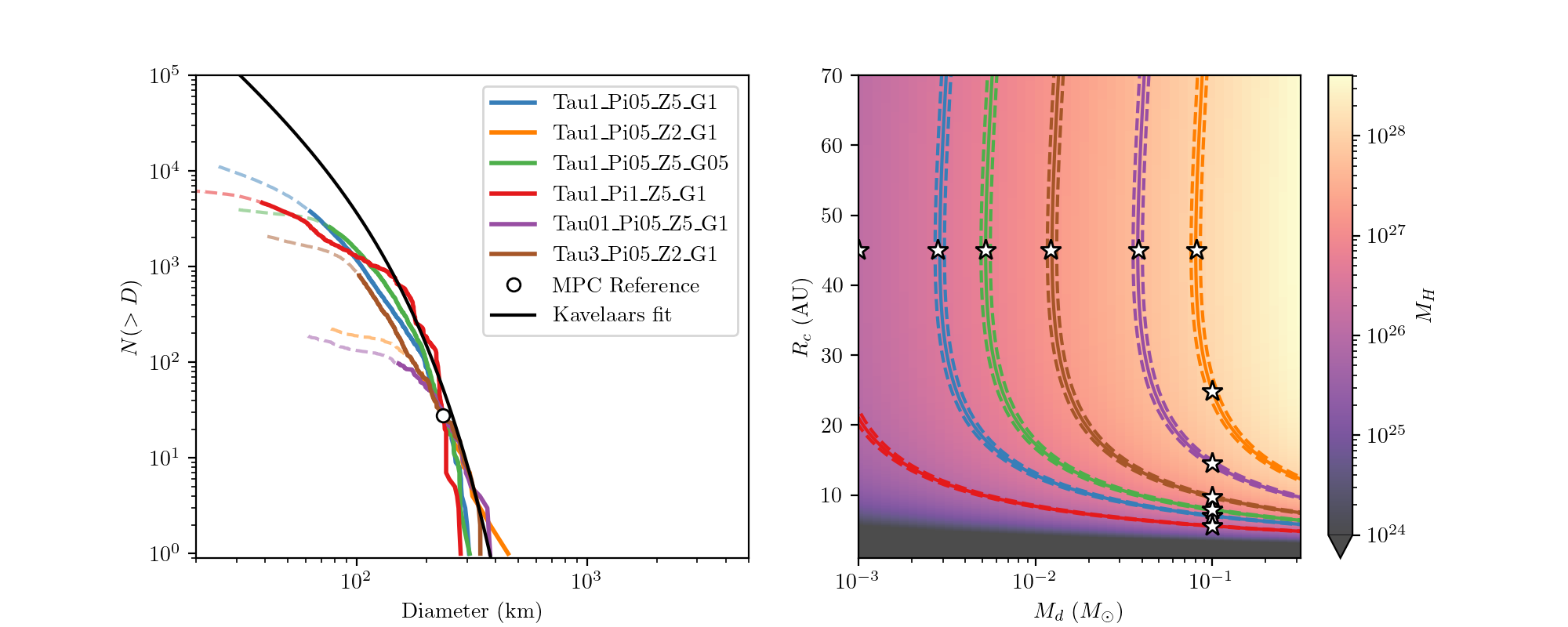}
    \caption{Parameter combinations that result in a $M_H$ that aligns the size distribution with the complete sample of the Minor Planet Center. Left: The scaled predicted size distributions, including objects at Hill radii both greater (solid) and less (dashed) than $2.5\times \Delta x$, along with the observational fit from \citealt{Kavelaars} (black curve) and the Minor Planet Center reference mass (see main text; white circle). Right: Plot of $M_H$ (color scale) for different combinations of $R_c$ and $M_d$. The overplotted curves show the $R_c$, $M_d$ combination that produces the $M_H$ value necessary to align the distributions of the same color shown in the left panel to observations. Here, $q$ varies between 0.4 and 0.6 (dashed lines), and $q = 0.5$ is shown as the solid lines. The values presented in Table~\ref{tab:model_values} are denoted by stars.}
    \label{fig:MH_scaling}
\end{figure}

\bibliography{bibliography}{}
\bibliographystyle{aasjournalv7.1}



\end{document}